\documentclass[reprint,amsmath,amssymb,pre,superscriptaddress,floatfix]{revtex4-2}
\usepackage{graphicx}
\usepackage{color}
\usepackage{dcolumn}
\usepackage{bm}
\usepackage{nicefrac}
\usepackage{hyperref}
\usepackage{siunitx}

\setcitestyle{super}

\renewcommand{\eqref}[1]{Eq.~(\ref{#1})}
\newcommand{\figref}[1]{Fig.~\ref{#1}}
\newcommand{\secref}[1]{Section~\ref{#1}}

\begin{document}
\title{A thermodynamic metric quantitatively predicts disordered protein partitioning and multicomponent phase behavior}

\author{Zhuang Liu}
\thanks{Authors contributed equally}
\affiliation{Department of Chemistry, Princeton University, Princeton, NJ 08544, USA}
\author{Beijia Yuan}
\thanks{Authors contributed equally}
\affiliation{Department of Chemistry, Princeton University, Princeton, NJ 08544, USA}
\affiliation{Lewis-Sigler Institute for Integrative Genomics, Princeton University, Princeton, NJ 08544, USA}
\affiliation{Department of Physics, Princeton University, Princeton, NJ 08544, USA}
\author{Mihir Rao}
\affiliation{Department of Chemistry, Princeton University, Princeton, NJ 08544, USA}
\affiliation{Department of Physics, Princeton University, Princeton, NJ 08544, USA}
\author{Gautam Reddy}
\affiliation{Department of Physics, Princeton University, Princeton, NJ 08544, USA}
\author{William M.~Jacobs}
\email{wjacobs@princeton.edu}
\affiliation{Department of Chemistry, Princeton University, Princeton, NJ 08544, USA}

\date{\today}

\begin{abstract}
  Intrinsically disordered regions (IDRs) of proteins mediate sequence-specific interactions underlying diverse cellular processes, including the formation of biomolecular condensates.
  However, quantitatively describing IDR interactions and phase behavior in complex multicomponent mixtures remains a major challenge for existing sequence-based models.
  Here, we introduce a thermodynamic model that predicts the free energies of arbitrary mixtures of IDRs across a wide range of concentrations.
  The model learns low-dimensional feature vectors for IDR sequences that combine linearly to define mixture representations from which free energies are predicted.
  Applying this framework to a coarse-grained simulation dataset of mixtures of IDR fragments, we show that the model predicts multicomponent phase diagrams in quantitative agreement with direct-coexistence simulations without being trained on free-energy or phase-coexistence data.
  We further show that the learned feature vectors define a thermodynamic metric space in which distances between IDRs quantify differences in their excess chemical potentials across mixture environments.
  This metric space provides geometrically intuitive predictions of IDR partitioning, multicomponent condensation, and context-dependent mutational effects, addressing several central problems in IDR biophysics within a single model.
  Systematic interrogation of the learned feature vectors reveals how amino-acid composition and sequence patterning jointly determine mixture thermodynamics.
  Together, our results establish a unified and interpretable framework for predicting and understanding the behavior of complex mixtures of IDRs and other sequence-dependent biomolecules.
\end{abstract}

\maketitle

\section{Introduction}

Interactions among intrinsically disordered regions (IDRs) of proteins influence cellular processes including transcriptional regulation, signaling, and the formation of biomolecular condensates.\cite{holehouse2024molecular,brangwynne2015polymer,shin2017liquid}
Despite being conformationally disordered, IDRs can drive the formation of complexes,\cite{borgia2018extreme} colocalize with specific partners,\cite{patil2023disordered} and selectively partition into distinct condensates.\cite{welles2024determinants,delacruz2024disorder,holehouse2025molecular}
These behaviors indicate that IDR interactions can be specific and that this specificity is encoded by the amino-acid sequences of the interacting IDRs.\cite{lyons2023functional,wessen2025sequence}

In addition to sequence dependence, IDR interactions can be \emph{context dependent}: the effective attraction or repulsion between two IDRs can depend not only on their identities, but also on the identities and concentrations of other components in a multicomponent mixture.
Context dependence arises naturally in mixtures of disordered biomolecules, whose conformational ensembles are influenced by their local environment.\cite{moses2020revealing,moses2024structural}
This motivates the development of predictive models that are not restricted to pairwise-additive descriptions with interaction coefficients that depend only on the identities of two IDRs; however, there is currently no practical framework for predicting context-dependent IDR interactions directly from sequences across mixtures with arbitrary compositions.
Moreover, because an IDR's interaction and partitioning preferences can vary with mixture composition,\cite{holehouse2025molecular,teilum2021specificity} a quantitative definition of interaction specificity must account for this context rather than rely solely on isolated pairwise binding measurements.

Aspects of these two problems have been addressed by various sequence-based approaches, which differ both in how they represent sequences and how these representations relate to interactions.
Supervised-learning models have been trained to classify phase-separation propensities of individual proteins,\cite{saar2021learning,chu2022prediction} predict whether pairs of proteins co-condense,\cite{zhang2024alphafold2} or categorize sequences by their partitioning behavior in selected condensates.\cite{saar2024protein,kilgore2025protein}
While these models can capture aspects of sequence and context dependence, they are optimized for specific prediction tasks and do not, by themselves, provide a general quantitative description of interactions and phase behavior in arbitrary multicomponent mixtures.
``Molecular grammar'' approaches instead represent sequences in terms of physically motivated amino-acid composition and patterning descriptors, providing mechanistic insights into sequence-dependent variations in IDR properties within individual protein families.\cite{wang2018molecular,martin2020valence,sun2024precise,zhang2025protein}
Grammar-based descriptors have been used to cluster IDRs via unsupervised learning, revealing relationships between sequence-based clusters and cellular functions.\cite{ruff2026molecular}
While these sequence representations provide a basis for measuring similarity, they do not, by themselves, establish a quantitative relationship between sequence similarity and context-dependent interaction preferences.

\begin{figure*}[t]
  \includegraphics[width=\textwidth]{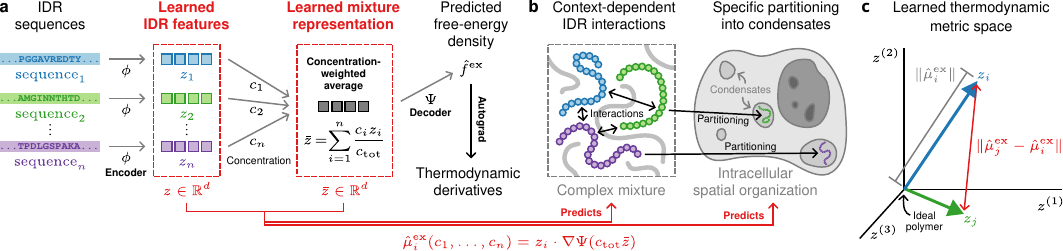}
  \caption{\textbf{An interpretable thermodynamic model predicts IDR interactions in multicomponent mixtures.}
    (a)~Context-independent IDR feature vectors combine in a concentration-weighted average to define a mixture representation, from which the excess free-energy density $\hat f^{\mathrm{ex}}$ is predicted.  These feature vectors are learned through joint optimization of the encoder and decoder.
    (b)~The learned feature vectors predict context-dependent interactions and phase behavior via the excess chemical potential function $\hat \mu^{\text{ex}}$, which is obtained by differentiating $\hat f^{\mathrm{ex}}$.
    (c)~These vectors reside in a thermodynamic metric space in which distances equal the norms of differences between $\hat \mu^{\mathrm{ex}}$ functions, $\|\hat\mu^{\mathrm{ex}}_j - \hat\mu^{\mathrm{ex}}_i\|$; since differences in $\hat \mu^{\mathrm{ex}}$ functions imply differences in context-dependent interactions, these distances quantify context-dependent interaction specificity.}
  \label{fig:1}
\end{figure*}

Alternatively, thermodynamic descriptions have been formulated in terms of mixture free energies, which can in principle be used to calculate various thermodynamic observables for any IDR mixture.
To this end, polymer-physics theories have been adapted to derive sequence-dependent free-energy models of IDRs in solution,\cite{lin2020unified,ghosh2022rules,li2022microphase,phillips2024beyond,wessen2022analytical,adachi2024predicting,ginell2025sequence} often using parameters obtained from experimentally validated coarse-grained force fields.\cite{dignon2018sequence,tesei2021accurate,joseph2021physics,saar2023theoretical,rizuan2026computational}
However, commonly used multicomponent free-energy models\cite{adachi2024predicting,ginell2025sequence} approximate interactions using composition-independent pairwise coefficients,\cite{colby2003polymer} which limits their ability to capture context dependence regardless of how their sequence-dependent parameters are determined.

Here we introduce a thermodynamic model that addresses both the prediction of sequence-encoded, context-dependent IDR interactions and the quantitative definition of interaction specificity in multicomponent mixtures.
First, the model predicts the free energies of arbitrary IDR mixtures across a wide range of concentrations without assuming pairwise additivity.
This is achieved by jointly learning representations of IDR sequences and a nonlinear function that maps their combined features to mixture free energies.
Second, these learned representations are embedded in a \textit{thermodynamic metric space}, in which Euclidean distances equal the norms of differences between excess chemical potential functions, providing a quantitative definition of context-dependent interaction specificity.
Together, these features of the model allow us to address several central questions in IDR biophysics---including the quantitative prediction of partitioning specificity, multicomponent condensation, and context-dependent mutational effects---within a unified framework.
We demonstrate these capabilities by training the model on simulation data generated using a coarse-grained force field and sequences derived from the human IDRome.
The model and its thermodynamic metric space thus provide a physically interpretable framework for predicting and explaining sequence-encoded, context-dependent interactions in complex mixtures.

\section{Results and Discussion}

\subsection{A thermodynamic metric for sequence space}

We hypothesized that the thermodynamic behavior of IDRs in multicomponent mixtures is governed by a small number of sequence-encoded molecular features.
Whereas prior analyses of low-dimensional thermodynamic behavior in biomolecular mixtures have relied on models with pairwise-additive free energies,\cite{graf2022thermodynamic,chen2024emergence,adachi2024predicting} we sought a description of IDR mixture thermodynamics that does not require this assumption.
We therefore developed a physics-based machine-learning framework that simultaneously identifies low-dimensional features of IDR sequences and learns how these features collectively determine the free energy of an IDR mixture.
This framework assumes that the homogeneous excess free energy is extensive and analytic in concentration, but does not assume pairwise additivity.
Crucially, the features identified via this framework are intentionally optimized to predict the free energies of IDR mixtures with arbitrary mole fractions across a wide range of concentrations.
Rather than being prescribed by prior theories of IDR physics, the learned features distill the sequence-encoded information required to predict the thermodynamic behavior of multicomponent IDR mixtures.

We represent each IDR by a $d$-dimensional \textit{feature vector}, $z$ (\figref{fig:1}a).
This learned representation of IDR $i$ is computed from the sequence by an \textit{encoder}, $\phi$, such that $z_i = \phi(\mathrm{sequence}_i)$; because this representation depends only on the sequence of IDR $i$, $z_i$ is independent of the mixture context.
Learned feature vectors combine via a concentration-weighted average to produce a \textit{mixture representation}, $\bar z = \sum_{i=1}^n (c_i/c_{\mathrm{tot}}) z_i$, where $c_1, \ldots, c_n$ are the concentrations of IDRs in the mixture and $c_{\mathrm{tot}} = \sum_{i=1}^n c_i$ is the total IDR concentration.
This linear mixing rule has several consequences that aid interpretability in the context of multicomponent mixtures, as we show throughout this paper.
The predicted excess free-energy density of a mixture is given by a nonlinear \textit{decoder} function, $\Psi$, such that $\hat f^{\mathrm{ex}} = \Psi(c_{\mathrm{tot}} \bar z)$.
Because the functional form of $\Psi$ is also learned, rather than prescribed by an approximate free-energy model, the model allows nonlinear mixture thermodynamics to be represented without assuming pairwise additivity.
We jointly optimize $\phi$ and $\Psi$ to predict mixture free energies, so that only thermodynamically relevant information is retained.
See Methods \secref{sec:model} and Supplementary Information (SI) Sections~I.A--G for complete details.

Within this framework, the predicted excess chemical potential function of IDR $i$, which describes the contribution of IDR $i$ to the free energy of an $n$-component mixture, is equal to
\begin{equation}
  \label{eq:muex}
  \hat\mu^{\text{ex}}_i(c_1, \ldots, c_n) = z_i \cdot \nabla \Psi(c_{\mathrm{tot}} \bar z),
\end{equation}
where $\nabla\Psi$ is the gradient of the learned decoder with respect to its argument.
\eqref{eq:muex} expresses a simple ``interaction rule'': context-dependent interactions arise from the alignment between a sequence feature vector and the composition-dependent gradient of the mixture free-energy landscape (\figref{fig:1}b).

We quantify the thermodynamic similarity of two IDRs by comparing their behavior across a distribution of mixture environments.
Given a mixture distribution $\rho_{\mathrm{mix}}$, we define the distance $\delta_{ij}$ between IDRs $i$ and $j$ by the norm of their \textit{exact} excess chemical potential difference
\begin{equation}
  \delta_{ij} \equiv \sqrt{\mathbb E_{c\sim\rho_{\mathrm{mix}}} \left[\left(\mu^{\mathrm{ex}}_i(c) - \mu^{\mathrm{ex}}_j(c)\right)^2\right]}.
\end{equation}
This establishes a \textit{thermodynamic metric space} in which IDRs $i$ and $j$ are close if and only if they behave similarly across mixtures drawn from $\rho_{\mathrm{mix}}$, independent of sequence similarity.
For a fixed $\rho_{\mathrm{mix}}$, the thermodynamic metric space is unique up to orthogonal transformations about the origin, which corresponds to an ideal molecule.

We then construct a physically meaningful coordinate system for the learned feature vectors such that
\begin{equation}
  \label{eq:distance}
  \|z_j - z_i\| = \hat \delta_{ij} \equiv \sqrt{\mathbb E_{c\sim\rho_{\mathrm{mix}}} \left[\left(\hat\mu^{\mathrm{ex}}_i(c) - \hat\mu^{\mathrm{ex}}_j(c)\right)^2\right]},
\end{equation}
which can be achieved by a linear transformation (see Methods \secref{sec:tm}).
Therefore, Euclidean distances $\|z_j - z_i\|$ in the learned thermodynamic metric space approximate $\delta_{ij}$ for sufficiently accurate predicted excess chemical potentials (\figref{fig:1}c).
We orient the coordinate axes using uncentered principal component analysis (PCA) of the transformed feature vectors.

\subsection{Low-dimensional thermodynamics of proteome-derived IDR mixtures}

\begin{figure*}[t]
  \includegraphics[width=\textwidth]{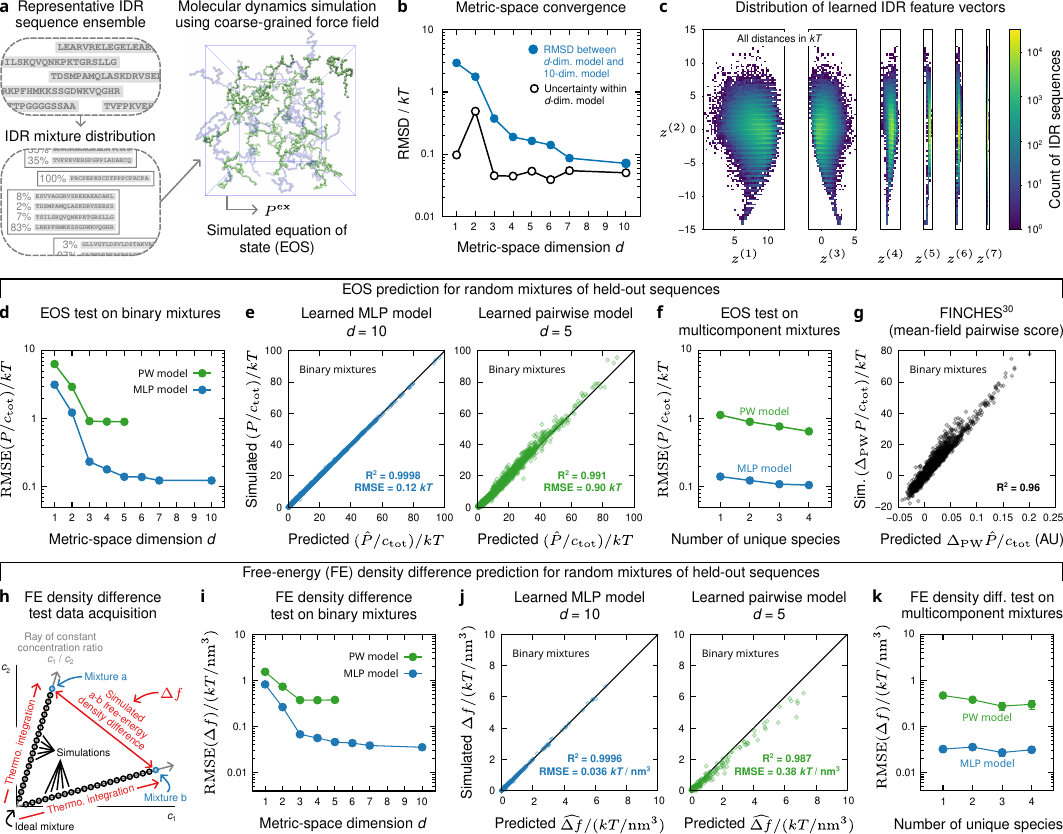}
  \caption{\textbf{Application of the framework to simulations of IDR fragments reveals a low-dimensional thermodynamic metric space and quantitatively predicts interactions in multicomponent mixtures.}
    (a)~We curated training and test IDRs by fragmenting the human IDRome into equal-length sequences of 20 residues.  This produced an ensemble of 335,439 minimally overlapping representative fragments, which were combined into mixtures by sampling from a mixture distribution.  We trained the thermodynamic model (cf.~\figref{fig:1}a) using equation of state (EOS) measurements obtained from coarse-grained simulations.
    (b)~Convergence of the MLP-model metric space with increasing dimension $d$.
    (c)~Two-dimensional projections of the learned feature vectors in the $d=10$ MLP-model metric space, with distances in units of $kT$.  Histograms show the distribution of all fragments in the sequence ensemble across the first 7 dimensions.
    (d)~EOS test performance, $\mathrm{RMSE}(P/c_{\mathrm{tot}})$, on 10,440 random binary mixtures as a function of the metric-space dimension $d$ and (e)~corresponding scatter plots for the MLP model with $d=10$ and a learned pairwise model with $d=5$.
    (f)~EOS test performance on random $n$-component mixtures for the same three models.
    (g)~Comparison of concentration-weighted pairwise scores predicted by FINCHES,\cite{ginell2025sequence} $\Delta_{\mathrm{PW}}\hat P$, with sequence-dependent pressure residuals extracted from EOS simulations, $\Delta_{\mathrm{PW}} P$ (see text).
    (h)~A test set of 231 free-energy density differences, $\Delta f$, between random binary mixtures was constructed by performing explicit thermodynamic integration.
    (i)~FE test performance, $\mathrm{RMSE}(\Delta f)$, as a function of the metric-space dimension $d$ and (j)~corresponding scatter plots for the MLP model with $d=10$ and a learned pairwise model with $d=5$.
    (k)~FE test performance, $\mathrm{RMSE}(\Delta f)$, on random $n$-component mixtures for the same two models.  Error bars indicate the standard deviation over random mixture pairings.
    In (d--k), sequences that compose the random mixtures were sampled uniformly from the metric space of the $d=10$ MLP model.}
  \label{fig:2}
\end{figure*}

In this paper, we show that this framework identifies low-dimensional IDR features that are sufficient to predict the free energy of IDR mixtures in coarse-grained simulations.
To define thermodynamic environments of biological interest, we constructed a statistically diverse sequence ensemble and mixture distribution.
We fragmented the human IDRome\cite{tesei2024conformational} into minimally overlapping segments of fixed length (20 residues), yielding 335,439 representative fragments (\figref{fig:2}a; see SI Sections~I.H--I).
This controls for leading-order polymer-length effects, allowing us to focus on sequence-dependent interactions, as in recent high-throughput experimental assays.\cite{kappel2025characterizing,norrild2026proteome}
We then constructed a mixture distribution by uniformly sampling sequences and biasing mole fractions toward mixtures enriched in a small subset of components, reflecting the selective recruitment characteristic of biomolecular condensates.
This mixture distribution defines the thermodynamic environments over which distances and specificity are evaluated.

We applied our framework to a well-validated, sequence-specific coarse-grained force field (Mpipi) that recapitulates experimentally observed behaviors across diverse IDR chemistries.\cite{joseph2021physics}
Because Mpipi is not tuned to a particular protein family or condensate type, it provides a suitable platform for uncovering broadly applicable sequence features.
We therefore used this force field to train an implementation of our framework that we refer to as the MLP model.
Rather than performing explicit free-energy or phase-coexistence calculations, we trained the model solely on equation-of-state (EOS) data obtained from inexpensive simulations (\figref{fig:2}a).
We then systematically varied $d$ to discover the dimensionality of the metric space.
See Methods \secref{sec:learning}, SI Sections~I.J--N, and Figs.~S1 and S2 for details.

We find that the thermodynamic behavior of mixtures of IDR fragments, as described by Mpipi, is intrinsically low dimensional: \textit{at most} $d \approx 10$ dimensions are sufficient to resolve differences in excess chemical potential functions across the mixture distribution to $\sim 0.1\,kT$ accuracy (\figref{fig:2}b).
Moreover, the first few dimensions account for most of the variance of feature vectors across the IDR sequence ensemble (\figref{fig:2}c).
This compact representation enables quantitative prediction of free energies, chemical potentials, and phase behavior in arbitrary mixtures without additional simulation.
We emphasize that the learned feature vectors and free-energy model presented in this paper pertain specifically to the selected sequence space and coarse-grained force field; nonetheless, our results demonstrate the ability of our framework to extract sequence features that describe IDR mixture thermodynamics and enable quantitative comparisons with existing theoretical predictions.

\subsection{Quantitative prediction of multicomponent mixture thermodynamics}

If the model learns how to accurately predict excess chemical potentials, it should quantitatively reproduce EOS data and free energies for mixtures not seen during training.
We therefore evaluated model predictions on mixtures composed exclusively of held-out IDRs.
For the $d=10$ MLP model, EOS predictions are in quantitative agreement with simulations of random binary mixtures sampled uniformly from the metric space (\figref{fig:2}d,e; see also Fig.~S3).
Errors decrease rapidly with increasing dimensionality and saturate near $d \approx 10$, mirroring convergence of the metric-space structure (\figref{fig:2}b).

To demonstrate the limitation of the pairwise-additive assumption, we trained a pairwise (PW) model by restricting the decoder to the form $\Psi^{\mathrm{PW}} = (\nicefrac{1}{2}) \sum_{i=1}^n \sum_{j=1}^n \varepsilon_{ij} c_i c_j + \hat f^{\mathrm{ref}}(c_{\mathrm{tot}})$, where $\hat f^{\mathrm{ref}}$ is a learned, monotonically increasing function.
The functional form of $\Psi^{\mathrm{PW}}$ is similar to that of the Flory--Huggins (FH) model,\cite{colby2003polymer} but here the rank-$d$ matrix of interaction coefficients $\varepsilon$ is predicted from sequences via an encoder--decoder architecture analogous to that used in the MLP model (see SI Section~I.E).
The PW model therefore provides a learned pairwise-additive baseline against which to assess the MLP model.
Despite being trained on the same data, the PW model resolves only approximately three thermodynamically meaningful dimensions, with prediction errors saturating at $\sim1\,kT$ (\figref{fig:2}d,e).
This indicates that the pairwise-additive approximation introduces significant bias in predicting the thermodynamics of IDR-fragment mixtures in Mpipi simulations.
At the same time, the pairwise-additive model learns a coarser representation of sequence-dependent interactions, resolving fewer thermodynamically meaningful metric-space dimensions than the MLP model.

Model performance improves as the number of distinct components in random mixtures increases from one to four (\figref{fig:2}f).
This improvement is consistent with self-averaging in multicomponent systems, facilitated by the linear mixing structure of the metric space.
Component-level variability partially averages out, enabling increasingly robust predictions as mixture complexity increases.

To benchmark our results against existing theoretical models constructed from the same coarse-grained force field, we compared our EOS simulation data with the predictions of FINCHES,\cite{ginell2025sequence} which estimates pairwise interaction strengths between IDRs using force-field-derived interactions supplemented by empirical corrections for local sequence chemistry.
To isolate sequence-dependent effects, we fit a monotonically increasing function $\hat P^{\mathrm{ref}}(c_{\mathrm{tot}})$ to the EOS simulation data and compared the residual of the simulation pressure $P$, $\Delta_{\mathrm{PW}} P \equiv P - \hat P^{\mathrm{ref}}(c_{\mathrm{tot}})$, to the FINCHES prediction for this residual, $\Delta_{\mathrm{PW}} \hat P = (\nicefrac{1}{2}) \sum_{i=1}^n \sum_{j=1}^n (\varepsilon_{ij}^{\mathrm{FINCHES}} - \bar\varepsilon^{\mathrm{FINCHES}}) c_i c_j$ (\figref{fig:2}g; see SI Section~I.O).
FINCHES predictions exhibit a good correlation with the simulation data, although weaker than those of the learned PW model.
Alternative theoretical models constructed from the Mpipi force field perform similarly (Fig.~S4).

Although the learned models were trained on EOS data, thermodynamic behavior is governed by the free-energy landscape.
To assess whether these models accurately predict free energies, we constructed a test set of 231 free-energy density differences between random binary mixtures by performing explicit thermodynamic integration of simulation data (\figref{fig:2}h).
Free-energy accuracy mirrors EOS accuracy overall (\figref{fig:2}i,j).
The MLP model accurately reconstructs the underlying free-energy landscape, while the errors incurred by the learned PW model are approximately an order of magnitude greater.
Free-energy accuracy remains stable with increasing component number (\figref{fig:2}k), demonstrating that the learned representation generalizes to increasingly complex multicomponent mixtures.

\subsection{Quantitative prediction of sequence-patterning effects at fixed amino-acid composition}

\begin{figure*}[t]
  \includegraphics[width=\textwidth]{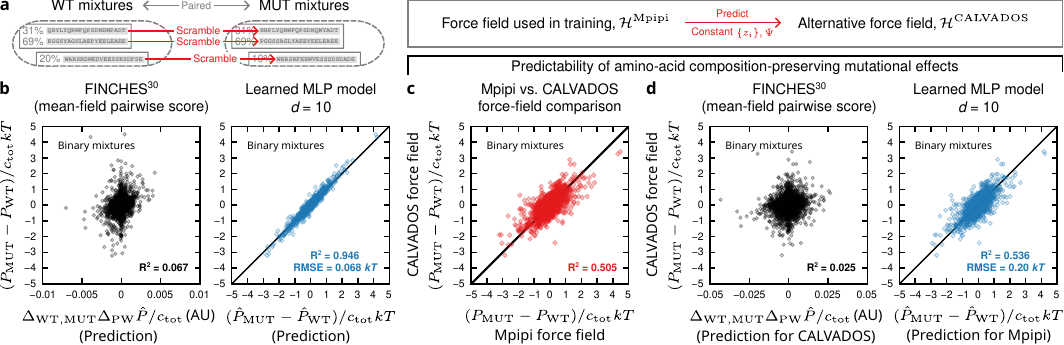}
  \caption{\textbf{The model predicts the thermodynamic effects of amino-acid composition-preserving mutations across force fields.}
    (a)~Paired wild-type (WT) and mutant (MUT) mixtures were constructed by replacing the test IDR sequences with scrambled sequences sampled according to ESM2 pseudo-likelihood,\cite{lin2023evolutionary} thus preserving their amino-acid compositions.
    (b)~Mutational effects were quantified by comparing EOS simulations of the WT binary-mixture test set (\figref{fig:2}e) with EOS simulations of MUT mixtures.  Predictions of the EOS difference between paired WT and MUT mixtures are shown for FINCHES, where $\Delta_{\mathrm{WT,MUT}}\Delta_{\mathrm{PW}}\hat P \equiv \Delta_{\mathrm{PW}} \hat P_{\mathrm{MUT}} - \Delta_{\mathrm{PW}} \hat P_{\mathrm{WT}}$, and the $d=10$ MLP model.
    (c)~Paired WT--MUT EOS differences measured in Mpipi and CALVADOS\cite{tesei2021accurate} simulations, which employ different force fields $\mathcal{H}^{\mathrm{Mpipi}}$ and $\mathcal{H}^{\mathrm{CALVADOS}}$, are strongly correlated across the same binary-mixture test set.
    (d)~Predictions of the EOS differences between paired WT and MUT mixtures in CALVADOS simulations are shown for FINCHES and the $d=10$ MLP model.  FINCHES predictions were derived directly from the CALVADOS force field, whereas the Mpipi-trained MLP model was evaluated as is, without retraining.
  }
  \label{fig:3}
\end{figure*}

We next asked whether the learned feature vectors accurately describe sequence patterning effects that are independent of the amino-acid composition.
Because no two IDRs in our ensemble share identical amino-acid composition, we sampled composition-preserving mutants by performing residue swaps and scoring candidate sequences with the ESM2 protein language model,\cite{lin2023evolutionary} favoring sequences consistent with learned protein-sequence statistics (see SI Section~I.I).

We performed EOS simulations for mutant mixtures by replacing each WT sequence in the binary-mixture test set with a MUT sequence (\figref{fig:3}a).
Comparison of paired WT and MUT simulations reveals composition-independent EOS differences between paired mixtures of up to $\pm 3\,kT$, with a standard deviation of $\sim 0.3\,kT$ (\figref{fig:3}b).
MLP-model predictions for MUT mixtures achieve the same level of accuracy as for mixtures of WT IDRs, demonstrating that the model generalizes to biologically plausible IDR fragments that are not present in the human IDRome.
By contrast, FINCHES predictions are minimally correlated with the simulated EOS differences.

To determine whether these predictions are transferable beyond the Mpipi force field, we repeated the EOS simulations of the same WT and MUT mixtures using the CALVADOS coarse-grained force field.\cite{tesei2021accurate}
Although Mpipi and CALVADOS are both residue-resolution coarse-grained force fields, they employ different functional forms and parameterizations of their nonbonded interactions.
Nevertheless, the thermodynamic effects of sequence patterning are strongly correlated between Mpipi and CALVADOS simulations (\figref{fig:3}c; see also Fig.~S5).
Consequently, the Mpipi-trained MLP model explains roughly half the variation in the CALVADOS simulation data without modification (\figref{fig:3}d).
FINCHES does not capture these sequence-patterning effects, despite utilizing the CALVADOS force field in the calculations.

Taken together, these results demonstrate that the MLP model learns subtle yet thermodynamically relevant details of IDR sequences.
The transferability of sequence-patterning effects between Mpipi and CALVADOS suggests that the learned features capture aspects of heteropolymer thermodynamics that are robust to differences in coarse-grained force fields.
Nevertheless, retraining or fine-tuning may be necessary to achieve comparable absolute accuracy when applying the framework to a different force field.

\subsection{Quantitative prediction of multicomponent phase diagrams}

\begin{figure*}[t]
  \includegraphics[width=\textwidth]{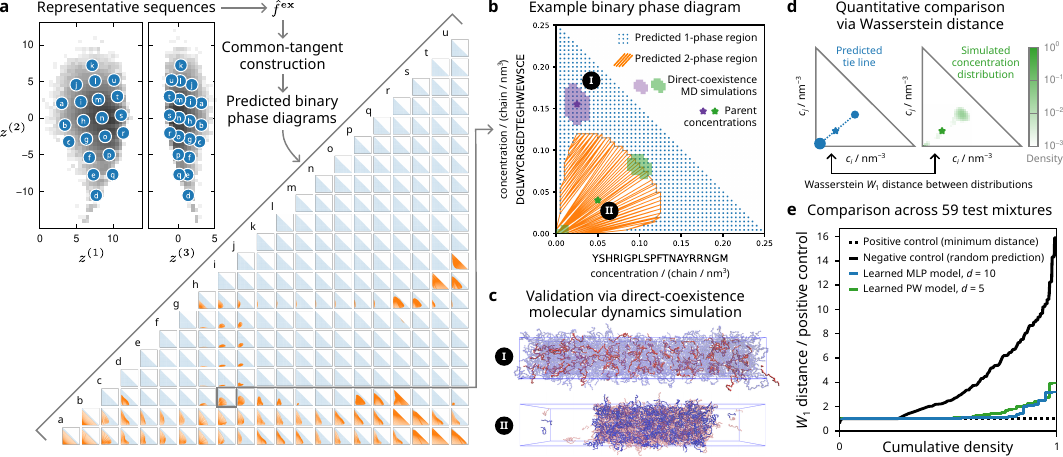}
  \caption{\textbf{The model quantitatively predicts multicomponent IDR phase diagrams.}
    (a)~Using representative sequences sampled uniformly from the metric space, we applied the $d = 10$ MLP model to predict a matrix of binary phase diagrams.
    (b)~In this example phase diagram (for representative sequences {\textsf c} and {\textsf i}), blue points indicate concentrations at which the mixture is predicted to be homogeneous.  Orange lines connect predicted coexisting phases within two-phase regions; mixtures with parent concentrations within these regions are predicted to phase-separate into coexisting phases connected by the orange tie lines.
    (c)~Predictions from (b) are compared with direct-coexistence molecular dynamics simulations at two selected parent concentrations, labeled I and II.  Equilibrium simulation snapshots indicate that parent concentration I is a homogeneous mixture, whereas parent concentration II phase separates.  The distributions of local concentrations observed in simulations are summarized by thresholded histograms (shaded regions) in (b).
    (d)~Comparisons are quantified using the Wasserstein $W_1$ distance between the predicted one- or two-phase distribution in the thermodynamic limit and the distribution of local concentrations in direct-coexistence simulations at the same parent concentration.
    (e)~The distribution of normalized $W_1$ distances was computed for a test set of 59 parent concentrations sampled from (a) for the learned MLP and PW models.  The positive control is the smallest possible $W_1$ distance for each mixture given the simulated concentration distribution.  The negative control was constructed by pairing simulations with random predictions that preserve the parent concentration.  Example phase diagrams for held-out sequences are shown in Fig.~S6.
  }
  \label{fig:4}
\end{figure*}

Having established that the model accurately predicts free-energy density differences, we next examined whether it captures the resulting phase behavior.
Because phase coexistence is determined by the global structure of the free-energy landscape,\cite{jacobs2017phase,jacobs2021self,zwicker2022evolved,chen2023programmable} it further tests whether the learned models and their associated thermodynamic metric spaces accurately describe collective behavior in the thermodynamic limit.
We therefore asked whether the model can quantitatively predict multicomponent phase diagrams without being trained on explicit examples of phase coexistence.\cite{jacobs2017phase,qian2024dominance}

We first generated binary phase diagrams for representative pairs of IDR sequences sampled uniformly from the metric space (\figref{fig:4}a).
For each pair, phase boundaries were computed directly from the predicted free-energy density using a multicomponent common-tangent construction,\cite{mao2019phase} without any training on coexistence data (see SI Section~I.P).
Each predicted phase diagram specifies whether a mixture at a given parent concentration remains homogeneous or phase-separates into coexisting phases connected by tie lines (\figref{fig:4}b).

\begin{figure*}[t]
  \includegraphics[width=0.67\textwidth]{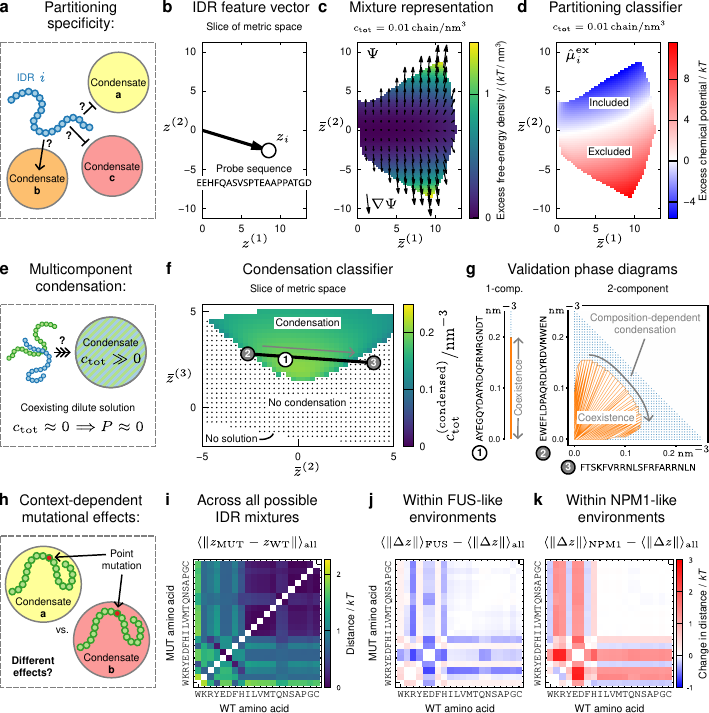}
  \caption{\textbf{The thermodynamic metric space organizes IDRs according to their physical interactions.}
    (a)~Partitioning specificity refers to the tendency of an IDR to be included in or excluded from condensates with different compositions.  This can be predicted directly from the geometry of the metric space, since the excess chemical potential of an IDR is the inner product of (b)~the IDR feature vector, $z_i$, and (c)~the gradient of the excess free-energy density, $\nabla\Psi$, evaluated at $c_{\mathrm{tot}} \bar z$.  (d)~This construction yields a classifier for preferential inclusion, $\hat\mu_i^{\mathrm{ex}}(c_{\mathrm{tot}} \bar z) < 0$, or exclusion, $\hat\mu_i^{\mathrm{ex}}(c_{\mathrm{tot}} \bar z) > 0$, of the probe IDR $i$ in (b) as a function of the mixture representation $\bar z$ in (c).  For visualization, we plot a two-dimensional slice of the metric space at constant $c_{\mathrm{tot}}$ in (c,d).
    (e)~Condensation occurs when a concentrated droplet coexists with a dilute solution.  (f)~This can be predicted directly within the metric space by looking for solutions to the condensation coexistence equation $P_{\mathrm{coex}} \approx 0$, yielding a classifier for both individual IDRs and IDR mixtures.  For visualization, we show a two-dimensional slice of the metric space.  The colored region indicates where $\hat P_{\mathrm{coex}} \approx 0$ has a nontrivial solution, leading to a droplet with the indicated $c_{\mathrm{tot}}$.  A single IDR condenses if its feature vector lies within this region, whereas an IDR mixture undergoes mole fraction-dependent condensation if the convex hull of its components' feature vectors intersects this region.  (g)~These predictions are confirmed by unary and binary phase diagrams, respectively, for probe IDR 1 and the pair of probe IDRs 2 and 3 shown in (f); in the binary case, moving along the line segment in (f) corresponds to sweeping through mole fractions from 100\% IDR 2 to 100\% IDR 3 in (g).
    (h)~The effects of point mutations on IDR interactions depend on the composition of the local environment.  (i)~The metric space can be exploited to quantify the effects of all single-amino-acid substitutions, $\|\Delta z\| = \|z_{\mathrm{MUT}}-z_{\mathrm{WT}}\|$.  These are averaged over the IDR sequence ensemble to obtain a matrix of mean mutation distances.  Because this calculation makes use of the uniformly sampled $\rho_{\mathrm{mix}}$ introduced in \figref{fig:2}a, it reflects mutational effects across diverse IDR mixtures.  Amino acids are ordered according to the leading wild-type mode of variation in the mutation-distance matrix, indicating an overall ranking of thermodynamic importance.  Specializing the distribution to mixtures composed only of (j)~FUS fragments, which are relatively hydrophobic, or (k)~NPM1 fragments, which are enriched in negative charge, rescales metric distances, revealing context dependence across distinct mixture distributions.
    All calculations were performed using the $d=10$ MLP model.
  }
  \label{fig:5}
\end{figure*}

We validated these predictions using direct-coexistence molecular dynamics simulations of 59 parent concentrations sampled from the representative phase diagrams in \figref{fig:4}a (see SI Section~I.Q).
Simulations either remained homogeneous or phase-separated into coexisting phases.
For each selected parent concentration, we compared the predicted phase behavior to the equilibrium distribution of local concentrations measured in a direct-coexistence simulation (\figref{fig:4}b,c).
Agreement was quantified using the Wasserstein (earth-mover) distance $W_1$ between predicted thermodynamic-limit concentration distributions and those obtained from finite-size simulations (\figref{fig:4}d).
Because direct-coexistence simulations are subject to significant finite-size effects arising from interfacial fluctuations and limited sampling volumes, we benchmarked $W_1$ against a positive control---the minimum $W_1$ distance achievable by a one- or two-phase thermodynamic-limit distribution consistent with the parent concentration---and a negative control constructed by pairing simulations with random predictions that preserve the parent concentration (\figref{fig:4}e).

Both the learned MLP and PW models perform near the positive control in the vast majority of test cases, indicating that discrepancies are comparable to those expected from finite-size effects in the direct-coexistence simulations.
Comparable results were obtained for random test mixtures assembled from sequences seen (\figref{fig:4}e) or held-out (Fig.~S6) during EOS training.

We also examined the correspondence between ternary phase diagrams predicted by the learned models and direct-coexistence simulations (Figs.~S7 and S8).
The number of test cases is limited by the extremely slow convergence of three-component direct-coexistence simulations.
Nevertheless, for two sequence triplets assembled from the representative sequences in \figref{fig:4}a, we find close agreement between equilibrated simulations and MLP predictions in both the number and compositions of phases at selected parent concentrations spanning one-, two-, and three-phase regions.
The limited simulation results are insufficient to resolve differences predicted by the MLP and PW models.

The strong agreement between predicted and simulated phase behavior demonstrates that the learned feature vectors are sufficient to reproduce multicomponent phase diagrams.
Moreover, phase diagrams can be rapidly predicted once the free-energy model is trained, whereas direct-coexistence simulations of multicomponent mixtures require extensive equilibration.
Together with EOS and free-energy validation, these results establish that the thermodynamic metric space captures sequence-dependent interaction specificity and provides quantitative predictions of phase behavior in the thermodynamic limit.
This quantitative foundation enables the direct geometric interpretation of the thermodynamic metric space that we turn to next.

\subsection{Geometric interpretation of partitioning, multicomponent condensation, and mutational effects}

The thermodynamic metric space organizes IDRs by their thermodynamic similarity rather than sequence similarity.
As a result, interaction specificity can be understood directly from geometric relationships in this space.
We illustrate this connection by answering three central questions in IDR thermodynamics: why specific IDRs are recruited into some condensates but excluded from others, why particular combinations of IDRs form droplets in coexistence with a dilute phase, and why the impact of a mutation depends on the ensemble of mixture environments that can be encountered.

Partitioning offers the clearest illustration of this geometric structure.
Relative to an ideal molecule, an IDR partitions into a mixture, such as one of the condensates illustrated in \figref{fig:5}a, when its excess chemical potential is negative, $\mu_i^{\mathrm{ex}} < 0$, and is excluded when $\mu_i^{\mathrm{ex}} > 0$.
Since the predicted excess chemical potential is given by an inner product between the feature vector $z_i$ and the gradient of the excess free-energy density, partitioning depends on their alignment evaluated at $c_{\mathrm{tot}} \bar z$.
Because both $z$ and $\bar z$ share the same metric space, partitioning reduces to a geometric classification problem (\figref{fig:5}b--d).
For a given IDR of interest and fixed total concentration, the metric space is divided into regions that either recruit or exclude that IDR as a function of the mixture representation $\bar z$ (\figref{fig:5}d).
Conversely, for a given mixture of interest, geometric classification identifies IDRs that are either included or excluded from that environment.
The thermodynamic geometry therefore provides immediate intuition: an IDR strongly partitions into a mixture when the gradient of the mixture's excess free-energy density is oppositely aligned with the IDR feature vector.

The same geometric framework also explains both single-component and multicomponent condensation.
Here we specifically consider a scenario in which an IDR-enriched droplet is in coexistence with a dilute solution, whose EOS satisfies $P \approx c_{\mathrm{tot}}^{(\mathrm{dil.})} kT \approx 0$ (\figref{fig:5}e; see SI Section~I.R).\cite{jacobs2021self}
Coexistence with a dilute solution occurs if the equation $\hat P(c_{\mathrm{tot}}^{(\mathrm{cond.})},\bar z) \approx 0$ admits a solution for the condensed-phase total concentration, $c_{\mathrm{tot}}^{(\mathrm{cond.})} > 0$, at fixed $\bar z$.
Thus, solving the heuristic equation $\hat P = 0$ allows us to partition the mixture $\bar z$ space into a condensation region where condensed mixtures can coexist with a dilute phase and a region in which such phase separation is not possible (\figref{fig:5}f).
A single-component IDR mixture can condense if its representation $\bar z = z$ lies within the condensation region.
For multicomponent mixtures, condensation is possible if the convex hull of component representations $\{z_1,\ldots,z_n\}$ intersects this region, leading to mole fraction-dependent condensation.
Unary and binary phase diagrams (\figref{fig:5}g) confirm these predictions for the example IDR sequences whose feature vectors are indicated in \figref{fig:5}f.
Thus, single-component and multicomponent condensation emerge from the same unified classifier in the metric space.
Although this particular classifier does not describe all possible phase-separation behaviors in a complex mixture and does not guarantee that a $\hat P \approx 0$ solution corresponds to a globally minimized free energy, it demonstrates how the learned thermodynamic geometry can be exploited to construct practically useful heuristics for multicomponent systems.

Beyond predicting mixture behavior, the metric structure also quantifies context-dependent mutational effects (\figref{fig:5}h).
The thermodynamic impact of a mutation is determined by the change in $\hat \mu^{\mathrm{ex}}$ across the mixture distribution, which is captured by the Euclidean distance between wild-type (WT) and mutant (MUT) feature vectors, $\|\Delta z\| = \|z_{\mathrm{MUT}} - z_{\mathrm{WT}}\|$.
Averaged over the IDR ensemble, this yields a matrix of mean point-mutation distances (\figref{fig:5}i; see SI Section~I.S).
Substitutions involving aromatic (W, Y, F) and charged (K, R, E, D) residues produce the largest overall perturbations, consistent with the dominant electrostatic, $\pi$--$\pi$, and cation-$\pi$ interaction modes encoded in the underlying force field.

Because distances depend on the mixture distribution, this framework naturally captures environmental specificity.
Restricting the mixture distribution to specific classes of IDR fragments, rather than sampling uniformly from the entire IDR-fragment ensemble, changes the relative magnitudes of mutational effects through a linear transformation of the feature vectors (Fig.~S9).
For example, restricting the distribution to mixtures of FUS IDR fragments (on average hydrophobic and weakly positively charged) or to mixtures of NPM1 IDR fragments (on average polar and strongly negatively charged) reveals that charged residues tend to exhibit the strongest differential effects in like-charged environments (\figref{fig:5}j,k).
Together, these examples illustrate how the mixture distribution enables both global and highly specialized calculations within a single model, without retraining.
The metric-space geometry therefore provides a unified and computationally efficient framework for predicting partitioning, condensation, and mutational sensitivity across diverse mixture contexts.

\subsection{Learned feature vectors encode amino-acid composition and sequence patterning}

\begin{figure*}[t]
  \includegraphics[width=0.67\textwidth]{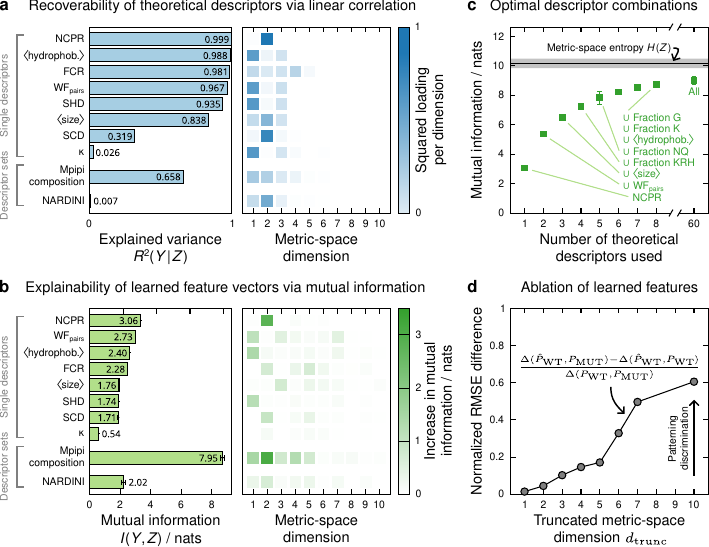}
  \caption{\textbf{Comparison of learned feature vectors with theoretical sequence descriptors distinguishes compositional and sequence-patterning determinants of IDR mixture thermodynamics.}
    (a)~The explained variance and corresponding metric-space loading vector, representing the optimal linear projection of theoretical descriptor(s) $Y$ onto learned feature vectors $Z$, are shown for each of eight individual descriptors, as well as for the 12-element Mpipi compositional descriptor set\cite{joseph2021physics} and the associated 40-element NARDINI patterning descriptor set.\cite{cohan2022uncovering}  Individual descriptors are net charge per residue (NCPR),\cite{mao2010net} average hydrophobicity,\cite{joseph2021physics} fraction of charged residues (FCR),\cite{mao2010net} a mean-field summary of short-ranged interactions ($\mathrm{WF}_{\mathrm{pairs}}$),\cite{von2025prediction} sequence hydropathy decoration (SHD),\cite{zheng2020hydropathy} average residue size,\cite{joseph2021physics} sequence charge decoration (SCD),\cite{firman2018sequence} and the charge-patterning descriptor $\kappa$.\cite{das2013conformations}
    (b)~The mutual information $I(Y,Z)$, which quantifies the statistical dependence, including nonlinear relationships, between $Y$ and $Z$, is shown along with the increase in $I(Y,Z)$ as each dimension is added to $Z$.
    (c)~Theoretical descriptor sets that best approximate the metric space, according to greedy maximization of $I(Y,Z)$.  The order in which the descriptors are added to the set is indicated; the final point uses all 60 descriptors, including Mpipi composition and NARDINI descriptors.  The entropy of the 335,439 IDR feature vectors, $H(Z)$, is shown for comparison (black line and gray region indicating SEM).
    (d)~Analysis of discrimination between WT and MUT mixtures (cf.~\figref{fig:3}b).  Binary-mixture EOS predictions using WT feature vectors, $\hat P_{\mathrm{WT}}$, are compared with simulations of WT, $P_{\mathrm{WT}}$, or MUT, $P_{\mathrm{MUT}}$, sequences; the normalized RMSE difference is given by the inset equation with $\Delta(A,B) \equiv \sqrt{\langle [(A - B) / c_{\mathrm{tot}}]^2 \rangle}$.  Ablation of learned feature-vector components greater than $d_{\mathrm{trunc}}$ indicates that higher-dimension components contribute to discrimination between shuffled sequences.
    All calculations were performed using the $d=10$ MLP model.  Error bars in (b,c) indicate the standard deviation of five independent estimates; errors are smaller than the symbol sizes in (d).
  }
  \label{fig:6}
\end{figure*}

\begin{figure*}[t]
  \includegraphics[width=\textwidth]{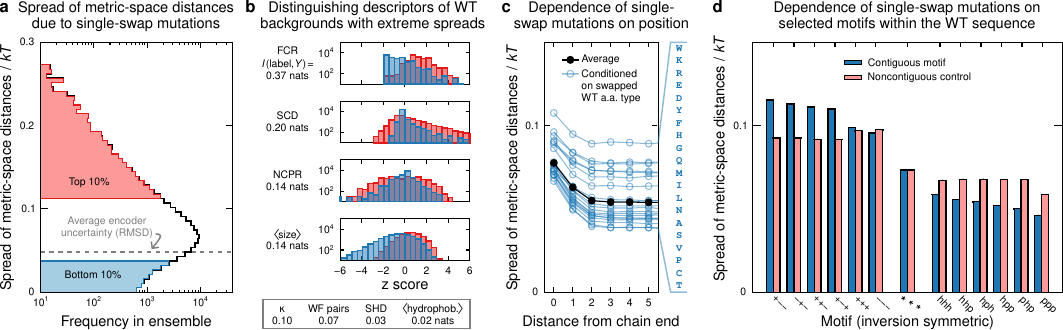}
  \caption{\textbf{Counterfactual analysis of single-swap mutants reveals patterning-dependent determinants of IDR mixture thermodynamics.}
    (a)~All unique single-swap mutants around each WT sequence were weighted according to ESM2 pseudo-likelihood.  The histogram shows the distribution of the spread (RMSD of feature vectors in the metric space) of these mutants around each of the 100,000 WT sequences in the test set.
    (b)~The distributions of selected theoretical descriptors for WT sequences with extreme spreads, lying in the bottom (B, blue) or top (T, red) 10\% of the spreads shown in (a).  The ability of each of the eight individual theoretical descriptors to predict the label T or B of an extreme WT sequence is quantified by the mutual information $I(\mathrm{label}, Y)$.
    (c)~The average spread conditioned on the distance of either swapped position from the chain end and the WT amino acids at the swapped positions; higher values indicate that counterfactuals satisfying these criteria result in larger perturbations on average.
    (d)~The average spread conditioned on whether the swapped positions affect a contiguous or non-contiguous motif within the WT sequence.  Results are presented for motifs consisting of triplets of charged ($+$ or $-$), hydrophobic/polar (h or p), or, as a control, all ($*$) amino acids.
    All calculations were performed using the $d=10$ MLP model.  Error bars, if shown, indicate SEM; otherwise, errors are smaller than the symbol sizes.
  }
  \label{fig:7}
\end{figure*}

Whereas the metric-space geometry enables interpretation of thermodynamic properties directly from learned sequence representations, this structure does not explain how the representations arise from sequences.
We therefore investigated the relationship between IDR sequences and learned feature vectors systematically.
To this end, we first asked whether previously developed sequence descriptors motivated by theoretical studies of IDR biophysics\cite{mao2010net,das2013conformations,firman2018sequence,zheng2020hydropathy,pal2024differential,von2025prediction,pei2026sequence} are reflected in the learned metric space and to what extent the learned metric space encodes additional thermodynamic information.

We related theoretical descriptors to the learned feature vectors of the $d=10$ MLP model using two complementary analyses of the sequence ensemble (see SI Sections~I.T--V).
Canonical correlation analysis (CCA) identifies the strongest \textit{linear} relationships between theoretical descriptors and the learned coordinates (\figref{fig:6}a).
This analysis shows that several theoretical descriptors associated with charge and short-range interactions are strongly aligned with the first few dimensions of the PCA-oriented metric space.\cite{agrawal2025charge}
In particular, net charge is largely captured by a single coordinate, $z^{(2)}$, consistent with the geometric interpretation of partitioning in \figref{fig:5}b--d, where mixtures with positive $\bar z^{(2)}$ attract IDRs with negative $z^{(2)}$ and vice versa.
Mutual information analysis\cite{song2019smile}, which quantifies \emph{arbitrary nonlinear} statistical dependencies, yields a similar ranking of features but reveals additional structure (\figref{fig:6}b and Fig.~S10).
As higher dimensions are included, the mutual information $I$ between theoretical descriptors and the learned coordinates continues to increase, revealing dependencies that are not apparent in the linear CCA analysis.
Notably, mutual information analysis indicates that the amino-acid composition\cite{joseph2021physics} and NARDINI patterning descriptor sets\cite{cohan2022uncovering} are more strongly related to the learned feature vectors than is apparent from their linear relationships, suggesting that compositional and patterning information is encoded nonlinearly in the learned metric space.

To assess how well combinations of theoretical descriptors approximate the metric space, we used a greedy procedure to select descriptor subsets that maximize $I$ and compared this to the entropy $H$ of the learned coordinates (\figref{fig:6}c; see SI Section~I.V).
Although theoretical descriptors capture substantial information about the learned coordinates, the estimated mutual information remains below the reference entropy of the learned representation, suggesting that these descriptors do not fully account for its structure.

\subsection{Sequence patterning information is concentrated in higher metric-space dimensions}

Since the learned feature vectors accurately describe sequence-patterning effects (\figref{fig:3}), we next asked where patterning information resides in the learned metric space.
Systematic ablation of higher-dimension coordinates of the $d=10$ learned metric space progressively reduces the ability to distinguish WT from MUT mixtures, with discrimination largely lost by $d_{\mathrm{trunc}} \approx 5$ (\figref{fig:6}d).
We therefore conclude that the first few uncentered PCA dimensions, which account for the greatest sequence-dependent $\hat\mu^{\mathrm{ex}}$ variations across the IDR-fragment ensemble (\figref{fig:2}c), primarily encode compositional determinants of thermodynamics, whereas higher dimensions capture additional information associated with sequence patterning.
This interpretation is consistent with the strongest association between compositional sequence descriptors and the first $\approx 5$ dimensions of the learned feature vectors (\figref{fig:6}b).

We emphasize that ablation of higher-dimension feature-vector components, which depends on the orientation of the metric space, is distinct from comparisons made across models trained with different feature-vector dimensions.
Separately trained lower-$d$ MLP models, as well as lower-$d$ learned PW models, can also capture sequence-patterning effects, albeit with reduced accuracy.
The preceding analyses therefore suggest that higher-$d$ MLP models achieve high accuracy in part by separating compositional and patterning information across different directions in the learned metric space.

\subsection{A systematic analysis reveals when and how patterning mutations matter}

Finally, we complemented the preceding analyses of patterning-information encoding with an exhaustive counterfactual analysis of patterning mutations that preserve amino-acid composition.
We generated all unique single-swap mutants, representing the minimal perturbations that preserve amino-acid composition, for each of the 100,000 WT sequences in the test set.
For each WT background, we quantified sensitivity to residue order by computing the spread of the mutant and WT feature vectors, defined as their ESM2-pseudo-likelihood-weighted root-mean-square deviation (RMSD) about the corresponding weighted centroid in the thermodynamic metric space (see SI Section~I.W).

Across this ensemble, single-swap-induced spreads typically exceed the uncertainty of the encoder (\figref{fig:7}a).
However, the magnitude of these effects varies substantially across WT sequence backgrounds.
To identify sequence properties associated with heightened sensitivity, we compared sequences in the top and bottom 10\% of the spread distribution.
We examined theoretical descriptor distributions for these two classes and quantified their informativeness using mutual information (\figref{fig:7}b).
Charge-related features are most predictive, led by the fraction of charged residues (FCR), which tends to be greater for high-sensitivity WT backgrounds.
By contrast, hydrophobicity-related features show similar distributions across the two classes.

We then used the counterfactual analysis to relate dominant patterning effects directly to the positions and amino-acid types involved in the single-swap mutations.
Two qualitative effects emerge clearly.
First, the model exhibits an absolute positional dependence along the chain.
Conditioning on swap position shows that swaps involving residues within two to three positions of either chain end produce substantially larger perturbations to feature vectors on average (\figref{fig:7}c and Fig.~S11a).
Stratification by amino-acid identity further reveals that strongly interacting residues (W, K, R, E, D, Y, F) generate the largest effects when located near chain ends.
This behavior is physically intuitive and consistent with mean-field theoretical analysis of chain-end effects,\cite{desouza2026polymer} suggesting that strongly interacting residues can form more intermolecular contacts when positioned near an end than when buried in the chain interior.

Second, we analyzed linear motifs defined as contiguous triplets of charged ($+$ or $-$) or hydrophobic/polar (h/p) residues.
For each motif class, we computed the average spread for swaps affecting such motifs and compared these to controls in which the same residue types were distributed throughout the sequence.
Charged and hydrophobic motifs exhibit distinct behaviors (\figref{fig:7}d and Fig.~S11b): rearrangements that alter adjacency between oppositely charged residues produce larger perturbations than controls, whereas clustered hydrophobic residues reduce the effect of swapping relative to controls.
However, mutations affecting like-charged blocks resemble controls, indicating that adjacency of opposite charges is the dominant effect uncovered by this motif analysis.
These opposing trends indicate that the thermodynamic consequences of local sequence clustering depend on the chemistry of the residues involved, rather than simply the degree of sequence blockiness.
Together, these analyses reveal how the thermodynamic effects of sequence patterning depend on amino-acid composition, residue position along the chain, and local sequence motifs, demonstrating that the learned metric space captures physically meaningful determinants of IDR interactions without prescribing them in advance.

\section{Conclusions}

We have introduced a machine-learning framework to isolate thermodynamically relevant features of IDR sequences.
These features enable quantitative prediction of IDR mixture free energies with context-dependent interactions through a learned free-energy function that does not assume pairwise additivity.
Using a sequence ensemble derived from the human IDRome and sequence-specific coarse-grained simulations, we show that the thermodynamic behavior of multicomponent IDR mixtures can be described by a surprisingly low-dimensional metric representation.
Sequence-dependent interaction specificity is encoded by the geometry of the thermodynamic metric space, enabling quantitative prediction of chemical potentials, mixture free energies, and multicomponent phase behavior directly from sequences and concentrations.

This framework is interpretable in two complementary senses.
From the perspective of sequence-based prediction, we can understand the thermodynamically relevant features of complex mixtures simply by averaging the features of their constituent IDRs, weighted by their mole fractions, regardless of whether these features reflect amino-acid composition alone or more complicated sequence-patterning information.
From the sequence-agnostic perspective of thermodynamic geometry, these same feature vectors have a precise physical interpretation: distances in the metric space equal the norms of differences between excess chemical potential functions across mixtures of interest.
Together, these complementary perspectives make predictions of partitioning specificity, single- and multicomponent condensation, and mutational sensitivity physically interpretable in terms of a common, sequence-encoded free-energy landscape.

Our results clarify several central aspects of IDR thermodynamics.
The collective behavior of IDRs emerges from a small number of sequence features that describe mixtures with varying numbers of components and context-dependent interactions.
Pairwise models can achieve partial accuracy when coefficients are learned, but their restricted functional form limits the thermodynamically relevant sequence information that can be resolved.
The influence of sequence patterning on mixture thermodynamics depends on the background amino-acid composition: patterning effects appear in higher dimensions of the metric space and are modulated by dominant compositional features encoded in lower dimensions.

Although the trained model of IDRome fragments that we report and analyze here pertains specifically to the selected sequence ensemble of IDR fragments and coarse-grained force field, the framework itself provides a general approach for identifying thermodynamically relevant representations of sequence-defined biomolecules and learning their mixture free energies.
The framework can be applied to alternative force fields and extended to other sequence spaces, including those containing sequences of different lengths.
Its structure also provides a practical route for systematic model refinement through the incorporation of multicomponent experimental data and simulations at mixed resolution.

Rather than treating the collective behavior of IDR mixtures as a collection of separate biophysical prediction problems, our thermodynamic metric approach derives diverse predictions from a single, sequence-dependent free-energy landscape.
In doing so, it establishes a general, interpretable foundation for understanding and predicting context-dependent interactions and phase behavior in biological and synthetic heteropolymers.

\begin{acknowledgments}
  This work is supported by the National Institute of General Medical Sciences of the National Institutes of Health under award number R35GM155017 to WMJ.
  Source code, pretrained models, and example calculations can be found at \texttt{github.com/wmjac/idr-frag-tm}.
\end{acknowledgments}

\section{Methods}

\subsection{A free-energy model for mixtures of sequence-defined biomolecules}
\label{sec:model}

We formulate IDR interactions at the level of mixture thermodynamics by predicting the free-energy density of a mixture as a function of sequence and concentration.
Once this quantity is known, all equilibrium thermodynamic observables---including the equation of state, chemical potentials, and phase behavior---follow by differentiation and free-energy minimization.
This formulation treats IDR interactions as emergent properties of the mixture rather than reducing them to pairwise affinities or abstract weights in classification models.

Our framework consists of an encoder--decoder architecture.
The encoder $\phi$ maps each IDR sequence to a context-independent feature vector $z \in \mathbb{R}^d$.
The mixture representation $\bar z \in \mathbb{R}^d$ is a concentration-weighted average of the feature vectors of its components, with context-dependent interactions emerging from the nonlinear response of the decoder to this collective representation.
The decoder $\Psi$ predicts the excess free-energy density relative to an ideal mixture, $\hat f^{\mathrm{ex}}$, from the decoder input $x = c_{\mathrm{tot}}\bar z$.
This formulation naturally enforces extensivity and permutation invariance,\cite{zaheer2017deep} yielding the following simple and general form for the excess free-energy density of an $n$-component mixture,
\begin{equation}
\label{eq:fex}
z_i = \phi(\mathrm{sequence}_i), \quad x = \sum_{j=1}^n c_j z_j, \quad \hat f^{\mathrm{ex}} = \Psi(x),
\end{equation}
with boundary conditions $\Psi(0) = \nabla \Psi(0) = 0$ that recover ideal-mixture behavior at vanishing concentration.
We implement the encoder using a transformer architecture to capture sequence dependence and the decoder using a multilayer perceptron (MLP).

For any finite ensemble of sequences, this formulation can approximate bulk free-energy densities to arbitrary accuracy in the limit of sufficiently large $d$ and sufficiently expressive encoder and decoder architectures.
Importantly, by learning from thermodynamic observables, a model with fixed $d$ acquires a representation optimized for mixture thermodynamics rather than predefined sequence descriptors.
Complete details are provided in SI Sections~I.A--E.

\subsection{Defining the thermodynamic metric space}
\label{sec:tm}

The excess chemical potential of IDR fragment $i$ is obtained by differentiating the learned excess free-energy density, $\hat \mu_i^{\mathrm{ex}} = \partial \hat f^{\mathrm{ex}} / \partial c_i$, yielding \eqref{eq:muex}.
\eqref{eq:muex} makes explicit that the thermodynamic properties of an IDR in an arbitrary mixture are determined by the alignment of its feature vector with the local gradient of the excess free-energy function, regardless of whether IDR $i$ has nonzero concentration in the mixture.

We define thermodynamic distances using the $L^2(\rho_{\mathrm{mix}})$ norm of $\hat \mu^{\mathrm{ex}}$ over a prior distribution of mixtures $\rho_{\mathrm{mix}}$.
We parameterize mixtures by total concentration $c_{\mathrm{tot}}$ and mole fractions $y_i \equiv c_i/c_{\mathrm{tot}}$, and assume a factorized prior $\rho_{\mathrm{mix}}(c_{\mathrm{tot}},y)=p(c_{\mathrm{tot}})\,p(y)$, with the mole fractions sampled from a sparse Dirichlet distribution.
We take $c_{\mathrm{tot}}$ to be uniform on $(0,\SI{0.25}{chain\per\nm\cubed}]$.
In the large-$n$ limit this yields sparse mixtures with an average effective cardinality ($\sim99\%$ mass) of 2.5, emphasizing environments enriched in only a few (typically 1--10 appreciably populated) components while sampling all possible sequence combinations.

We independently trained $K=5$ models using five-fold cross-validation, yielding learned latent vectors $\{\tilde z_i^{(k)}\}$ and a corresponding decoder $\tilde\Psi^{(k)}$ for each trained model $k = 1, \ldots, K$.
Following from \eqref{eq:muex}, the mixture prior induces a positive semidefinite matrix for each model $k$,
\begin{equation}
  M^{(k)}=\mathbb E_{c\sim\rho_{\mathrm{mix}}}\left[\nabla\tilde\Psi^{(k)}(\tilde x^{(k)}(c))\nabla\tilde\Psi^{(k)}(\tilde x^{(k)}(c))^{\mathsf T}\right],
\end{equation}
which we estimate by Monte Carlo sampling from $\rho_{\mathrm{mix}}$.
The squared $L^2(\rho_{\mathrm{mix}})$ norm of $\hat \mu_i^{\mathrm{ex}}$ is therefore $\tilde z_i^{(k)} M^{(k)} (\tilde z_i^{(k)})^{\mathsf T}$, defining an inner product on the latent space.
Defining $A^{(k)}=(M^{(k)})^{1/2}$ as the principal matrix square root yields the linear transformation
\begin{equation}
  \label{eq:transform}
  z_i^{(k)} = \tilde z_i^{(k)} (A^{(k)})^{\mathsf T},
\end{equation}
so that Euclidean distances in the transformed coordinates reproduce the $\hat \mu^{\mathrm{ex}}$ norm.
Importantly, because \eqref{eq:transform} is linear, the mixture prior can be changed to redefine distances without retraining the model, as demonstrated in \figref{fig:5}j,k.
This procedure converts each learned latent space into a consistent thermodynamic metric space in which distances satisfy \eqref{eq:distance}, with a corresponding transformed decoder $\Psi^{(k)}$.

After applying the metric transformation, we canonicalize each of the $K$ models by uncentered principal component analysis, fixing axis orientations by a consistent sign convention.
This procedure aligns the $K$ models in a common coordinate system.
Each canonicalized model defines a thermodynamic metric space in which Euclidean distances exactly reproduce differences in predicted excess chemical potential functions.
We report ensemble-averaged feature vectors to summarize the learned representations and quantify their uncertainty by the RMSD across the $K$ models.
See SI Sections~I.F--G for further details.

\subsection{Learning free energies from molecular simulations}
\label{sec:learning}

Free energies are not direct observables in molecular simulations.
Rather, they are typically obtained through thermodynamic integration,\cite{frenkel2001understanding} which is computationally expensive.
To circumvent this limitation, we trained the model on a thermodynamic derivative that is directly accessible: the excess contribution to the pressure, $P^{\mathrm{ex}}$, defined as the deviation of the pressure from ideal-mixture behavior.
The excess pressure can be computed straightforwardly from standard molecular dynamics simulations in the canonical (NVT) ensemble; we refer to these simulations in which $P^{\mathrm{ex}}$ is measured as equation-of-state (EOS) simulations.
Unlike phase-coexistence calculations, which may or may not show phase separation, each EOS simulation yields a well-defined pressure measurement so that all simulations contribute information during training.

The predicted excess pressure, $\hat P^{\mathrm{ex}}$, is obtained by automatic differentiation of the learned free-energy density,
\begin{equation}
  \hat P^{\mathrm{ex}}=x\cdot\nabla\Psi(x)-\Psi(x),
\end{equation}
ensuring thermodynamic consistency by construction.
The boundary conditions eliminate the linear ambiguity in reconstructing excess free energies from pressure data, ensuring that the predicted excess chemical potentials are uniquely defined.
We therefore trained the model by minimizing the mean squared error (MSE) of $P^{\mathrm{ex}} / c_{\mathrm{tot}}$, whose simulation uncertainties are approximately independent of concentration.
As in conventional thermodynamic integration, this EOS training strategy allows the use of simulations comprising a relatively small number of chains without incurring substantial finite-size effects.
Moreover, EOS data from thermodynamically unstable mixtures can be included during training, since equilibrium phase behavior is determined by the convex envelope of the learned free-energy landscape rather than its nonconvex regions.
This approach also avoids training directly on phase-coexistence simulations, which are computationally expensive and can be sensitive to equilibration and finite-size effects.

IDR fragments were divided into training, validation, and test sets, with test mixtures constructed exclusively from held-out sequences.
Training data were generated from simulations of random unary and binary mixtures of IDR fragments.
The model was trained end-to-end, jointly learning the sequence encoder and mixture decoder so that similarities among IDR sequences could be exploited to reduce data requirements.
See SI Sections~I.H--N for further details.

\end{document}


\setcounter{page}{28}
\setcounter{section}{3}
\section{Supplemental Figures}

\makeatletter
\begingroup
\@starttoc{lof}
\endgroup
\makeatother

\clearpage

\renewcommand{\thefigure}{S\arabic{figure}}


\begin{figure}[p]
  \centering
  \includegraphics[width=\linewidth]{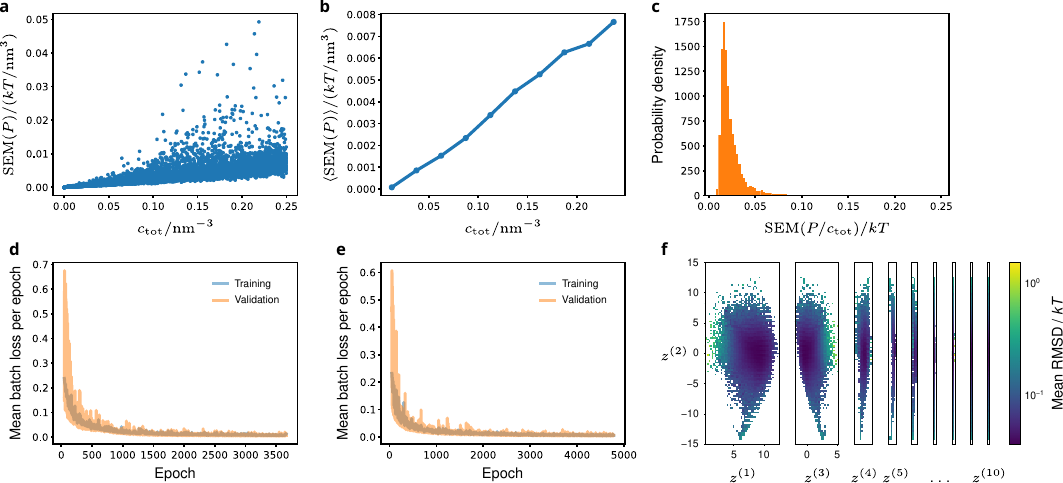}
  \caption[EOS simulation uncertainties and MLP model training.]
          {\textbf{Uncertainties in EOS simulations, MLP-model training curves, and metric-space uncertainty.}
            (a)~Standard error of the mean (SEM) of pressure values in the uniform-metric-space EOS test dataset.
            (b)~Mean EOS SEM of the data shown in (a), binned according to the total concentration, $c_{\mathrm{tot}}$.
            (c)~Histogram of the EOS SEM normalized by $c_{\mathrm{tot}}$, demonstrating that the simulation uncertainty on $P/c_{\mathrm{tot}}$ is roughly constant across concentrations and mixture compositions.  Comparison with the $d=10$ MLP-model loss on this test dataset (Fig.~2d,e) shows that the model achieves an accuracy within a factor of $\sim3$ of the minimum theoretical loss given by $\sim \sqrt{2} \times \langle \mathrm{SEM}(P/c_{\mathrm{tot}}) \rangle$.
            (d,e)~Mean mini-batch loss per epoch for the MLP model with metric-space dimension $d=10$ on the training and validation datasets.  Two example splits from five-fold cross-validation are shown, each using a randomly selected seed.  The loss decreases and stabilizes for both, with minimal evidence of overfitting.
            (f)~Projections of the mean encoder uncertainty across the metric space of the $d=10$ MLP model (cf.~Fig.~2c).
            }
  \label{fig:S1}
\end{figure}
\clearpage

\begin{figure}[p]
  \centering
  \includegraphics[width=\linewidth]{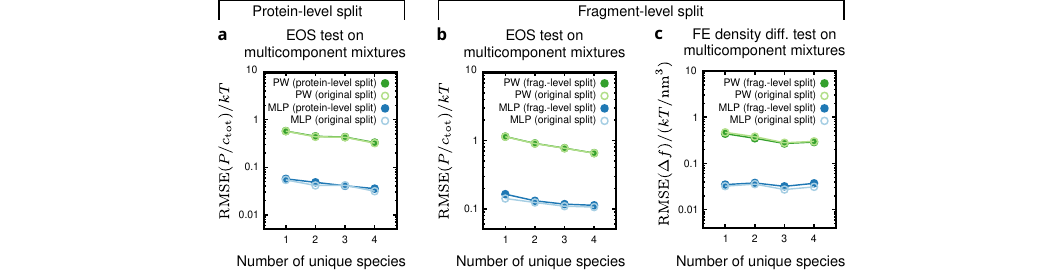}
  \caption[Effect of alternative training/test splits on model performance.]
          {\textbf{EOS and free-energy density difference test performance is consistent across alternative training/test splits.}
            Test performance is shown for both the learned PW and MLP models as a function of the number of unique components, $n$, in the test mixtures.
            (a)~EOS test performance using a protein-level split, in which 2,971 of the 15,347 human proteins were randomly reserved for testing and all training samples containing fragments from these proteins were removed. Models retrained on this dataset (protein-level split) are evaluated on new EOS test sets comprising 1,000 random mixtures for each $n=1,\ldots,4$, with sequences drawn exclusively from the held-out proteins. Predictions from the original models (original split) on these same test datasets are shown for comparison.
            (b)~EOS test performance using a fragment-level split, in which training samples were filtered to ensure that fragments originating from the same parent IDR and sharing 10 contiguous residues were not split between the training and test sets. The performance of each model retrained on this filtered dataset (fragment-level split) is compared with that of the corresponding original model (original split) on the original test datasets (cf.~Fig.~2f).
            (c)~Free-energy density difference test performance for the fragment-level split, comparing retrained and original models on the original test datasets (cf.~Fig.~2k).
          }
  \label{fig:S2}
\end{figure}
\clearpage

\begin{figure}[p]
  \centering
  \includegraphics[width=\linewidth]{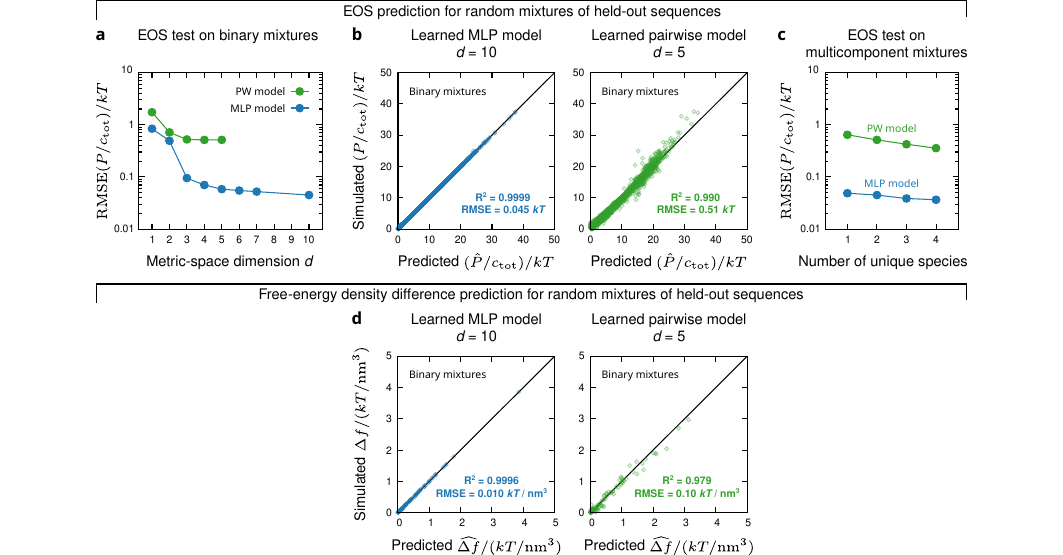}
  \caption[Test performance on a uniform-sequence-space test set.]
          {\textbf{EOS and free-energy density difference test performance for random mixtures composed of sequences sampled uniformly from the sequence ensemble.}
            All calculations are analogous to those shown in Fig.~2d--k.
            Overall, test performance tends to be better for the learned models on these datasets relative to the uniform-metric-space datasets presented in Fig.~2\mbox{d--k}, since sequences sampled uniformly from the sequence ensemble tend to be concentrated near the center of the metric-space distribution (see Fig.~2c), where the encoder uncertainty tends to be lowest (see Fig.~S1f).
            (a)~EOS test performance evaluated on 7,140 random binary mixtures as a function of the metric-space dimension $d$.
            (b)~EOS scatter plots for the $d=5$ learned pairwise model and the $d=10$ MLP model, evaluated on random binary mixtures.
            (c)~EOS test performance for the learned pairwise and MLP models as a function of the number of unique components in a multicomponent mixture.
            (d)~Free-energy density difference test performance evaluated on 100 random binary mixtures for the learned pairwise and MLP models.
          }
  \label{fig:S3}
\end{figure}
\clearpage

\begin{figure}[p]
  \centering
  \includegraphics[width=\linewidth]{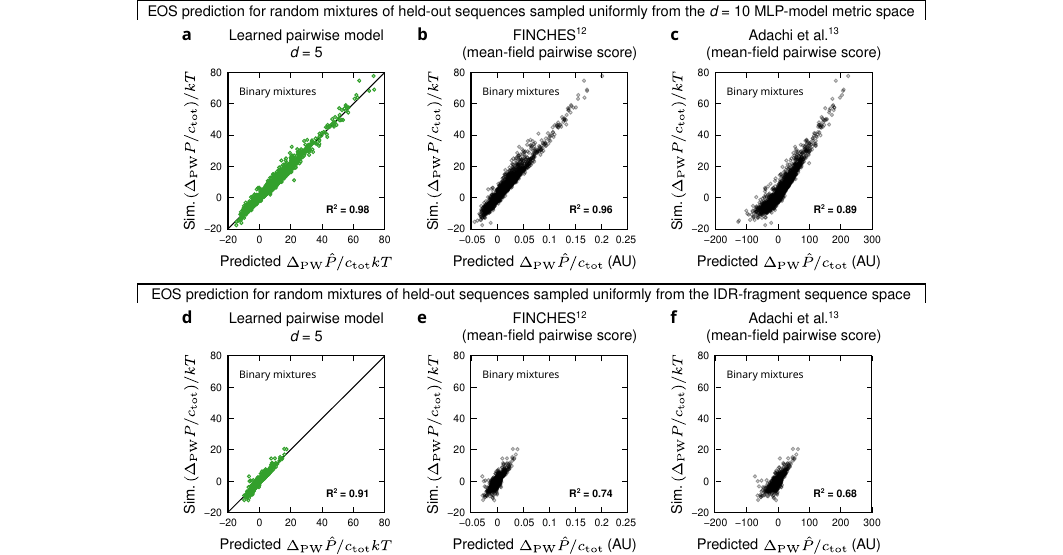}
  \caption[Comparison of pairwise theories.]
          {\textbf{Comparison of EOS prediction performance using learned and mean-field pairwise theories.}
            Comparisons of pairwise-model predictions with binary-mixture EOS simulation data are shown in the same format as Fig.~2g.
            Simulation data were used to compute $\Delta_{\mathrm{PW}} P \equiv P - \hat P^{\mathrm{ref}}(c_{\mathrm{tot}})$, where $\hat P^{\mathrm{ref}}(c_{\mathrm{tot}})$ is a fitted, monotonically increasing function of total concentration alone.
            These residuals were predicted using the pairwise coefficients obtained from the FINCHES and Adachi et al. mean-field models using the same force field, as described in Sec.~I.O.
            We also include comparisons with the learned pairwise model, in which case the predicted residuals are $\Delta_{\mathrm{PW}} \hat P = (\nicefrac{1}{2}) \sum_{i=1}^n \sum_{j=1}^n \varepsilon_{ij} c_i c_j-(\nicefrac{1}{2})\bar\varepsilon c_{\mathrm{tot}}^2$, with learned coefficients $\varepsilon_{ij}$, as described in Sec.~I.E.
            (a--c)~Comparisons of simulated and predicted pairwise residuals on the uniform-metric-space binary-mixture test dataset (cf.~Fig.~2d,e) for all three pairwise models; panel b is identical to Fig.~2g.
            (d--f)~Comparisons of simulated and predicted pairwise residuals on the uniform-sequence-space binary-mixture test dataset (cf.~Fig.~S3a,b) for all three pairwise models.
            }
  \label{fig:S4}
\end{figure}
\clearpage

\begin{figure}[p]
  \centering
  \includegraphics[width=\linewidth]{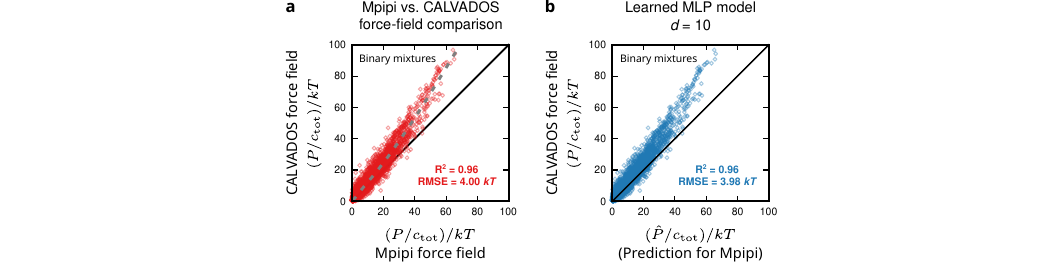}
  \caption[Comparison of Mpipi and CALVADOS force fields.]
          {\textbf{Comparison of Mpipi and CALVADOS force fields.}
            (a)~Comparison of EOS measurements between Mpipi and CALVADOS simulations using the WT binary-mixture test set (see Fig.~2d,e).  The gray dashed line represents a monotonic function $\hat g$ chosen to minimize the squared residuals of $P^{\mathrm{CALVADOS}}/c_{\mathrm{tot}}kT - \hat g(P^{\mathrm{Mpipi}}/c_{\mathrm{tot}}kT)$.
            (b)~Predictions of EOS measurements in CALVADOS simulations, using the same WT binary-mixture test set, are shown for the $d=10$ MLP model, which was evaluated without retraining.  Because the MLP model was trained on the Mpipi force field and the RMSE between the force fields in panel~a is $\sim 4 kT$, the RMSE between the MLP predictions and the CALVADOS EOS measurements is also $\sim 4 kT$.  Utilizing the mixture-independent transfer function $\hat g$ to define nonlinearly rescaled CALVADOS predictions $\hat P^{\mathrm{CALVADOS}} / c_{\mathrm{tot}}kT = \hat g(\hat P^{\mathrm{Mpipi}} / c_{\mathrm{tot}}kT)$ reduces the RMSE to $2.22 kT$.  Further improvement would require retraining or fine-tuning the MLP model.}
  \label{fig:S5}
\end{figure}
\clearpage

\begin{figure}[p]
  \centering
  \includegraphics[width=\linewidth]{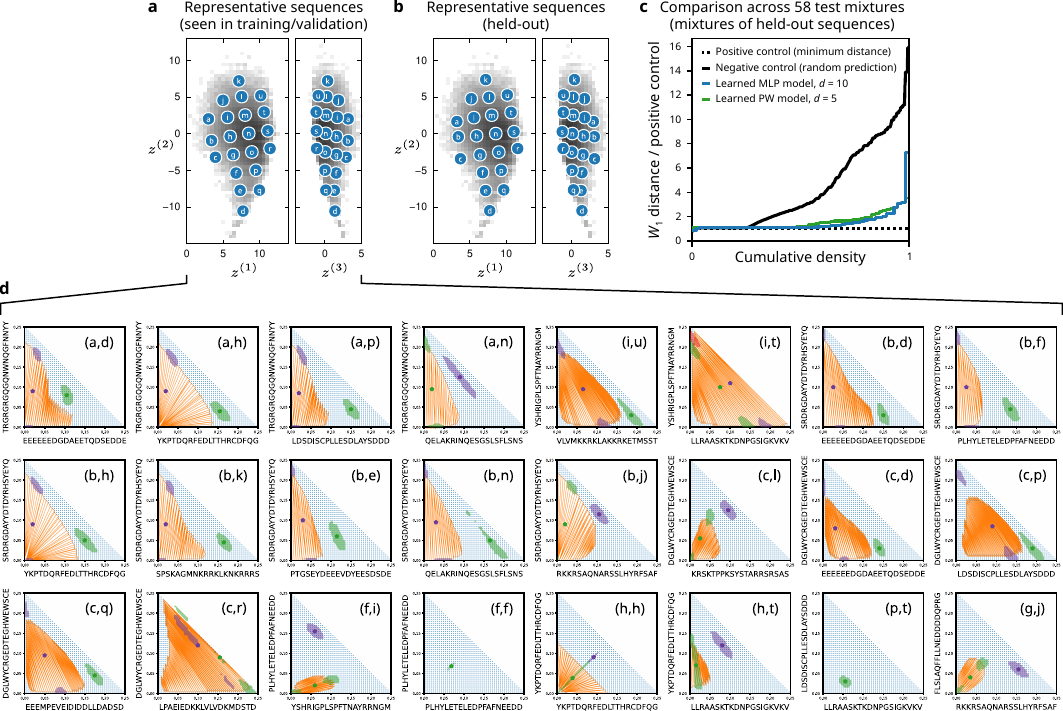}
  \caption[Binary phase diagrams for random mixtures of seen and unseen sequences.]
          {\textbf{Comparison of predicted phase diagrams and simulated phase coexistence for random binary mixtures of representative sequences.}
            (a)~Representative sequences seen in training and validation, sampled uniformly from the metric space (reproduced from Fig.~4a).
            (b)~Representative held-out sequences, sampled uniformly from the metric space.
            (c)~Comparison of predicted and simulated phase coexistence for random binary mixtures of held-out sequences, following the same format as Fig.~4e.  The error in the tail of the cumulative density for the MLP model is due entirely to phase diagrams involving sequence {\textsf b}, which has an anomalously high encoder uncertainty (see Fig.~S1f).  Otherwise, the predictive performance of the models is similar to that shown in Fig.~4e for mixtures of representative sequences seen in training and validation, which tend to have lower encoder uncertainty.
            (d)~Example predicted binary phase diagrams and simulation comparisons from the test set shown in Fig.~4e.  Units and symbols are as in Fig.~4b.  Tuples indicate the coordinates of each phase diagram in Fig.~4a.  Three-phase coexistence is neither predicted nor observed for binary mixtures of the sequence pairs shown in panel~a.}
  \label{fig:S6}
\end{figure}
\clearpage

\begin{figure}[p]
  \centering
  \includegraphics[width=\linewidth]{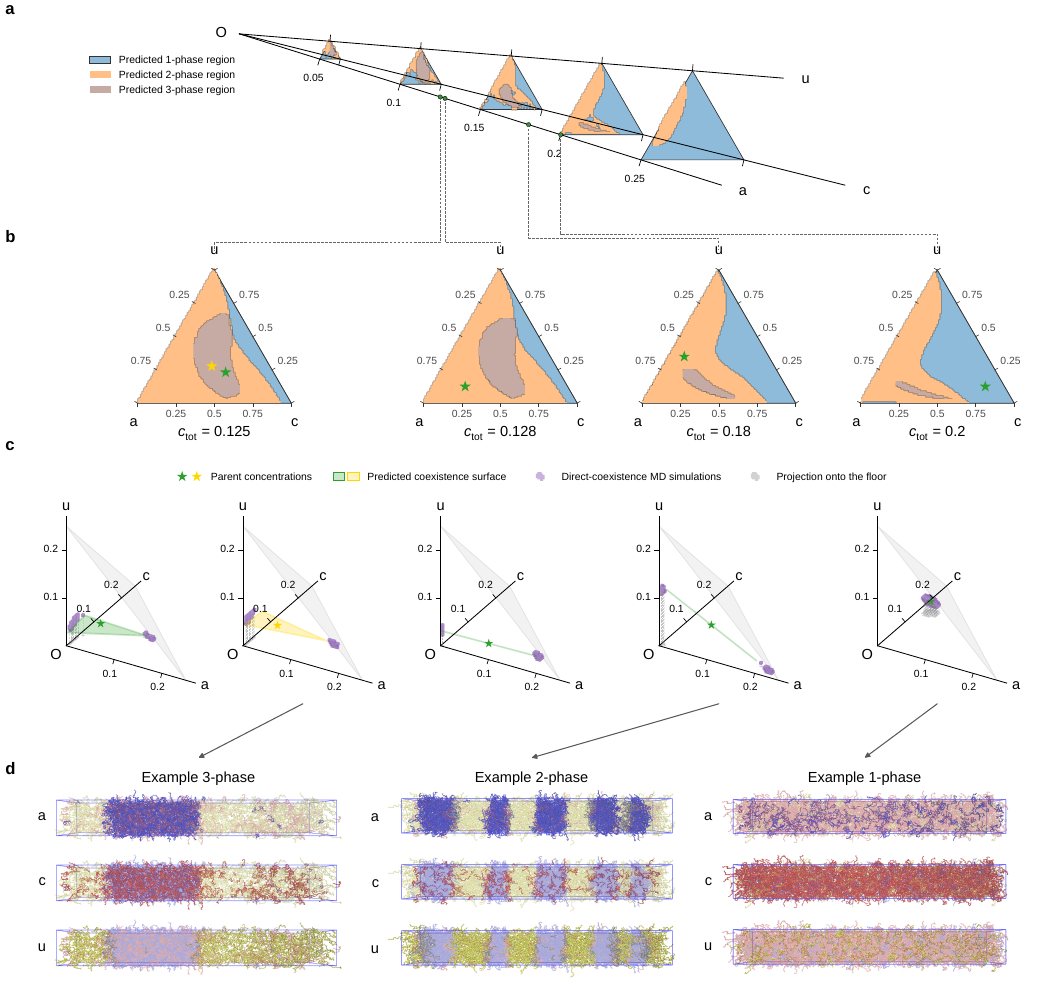}
  \caption[Example ternary phase diagram exhibiting two- and three-phase coexistence.]
          {\textbf{Ternary phase diagram for mixtures of representative sequences {\textsf a}, {\textsf c}, and~{\textsf u}.}
            (a)~Ternary phase diagram predicted by the $d=10$ MLP model.  We constructed the lower convex hull of the three-dimensional free-energy density surface $f(c_{\textsf a},c_{\textsf c},c_{\textsf u})$ and classified each grid point as homogeneous or as separating into two or three coexisting phases.  Slices of the ternary phase diagram are shown at $c_{\mathrm{tot}}=0.05$, $0.1$, $0.15$, $0.2$, and $\SI{0.25}{\chain\per\nm\cubed}$.
            (b)~Parent concentration vectors selected for coexistence simulations are indicated by green or yellow stars.  Two selected mixtures are predicted to form three phases, two are predicted to form two phases, and one is predicted to remain homogeneous.
            (c)~Predicted coexisting phases in the three-dimensional concentration space, with projections onto the $c_{\textsf u}=0$ plane to aid depth perception.  The predicted tie triangles and tie lines (green or yellow) are overlaid on local-concentration histograms from simulations (purple).  Simulations agree well with predictions for parent mixtures predicted to be homogeneous or to form two phases.  For mixtures within the predicted three-phase region, the simulated local-concentration distributions encompass the compositions of all three predicted coexisting phases.  The simulated local-concentration distribution spans the predicted tie line connecting two compositionally similar phases, although their similar compositions make them~difficult}
  \label{fig:S7}
\end{figure}
\clearpage

\noindent{\small to resolve as separate clusters in finite-size simulations.
  (d)~Snapshots from direct-coexistence simulations of selected mixtures.  In each row, one species is shown opaque.}
\vskip5ex

\begin{figure}[h]
  \centering
  \includegraphics[width=\linewidth]{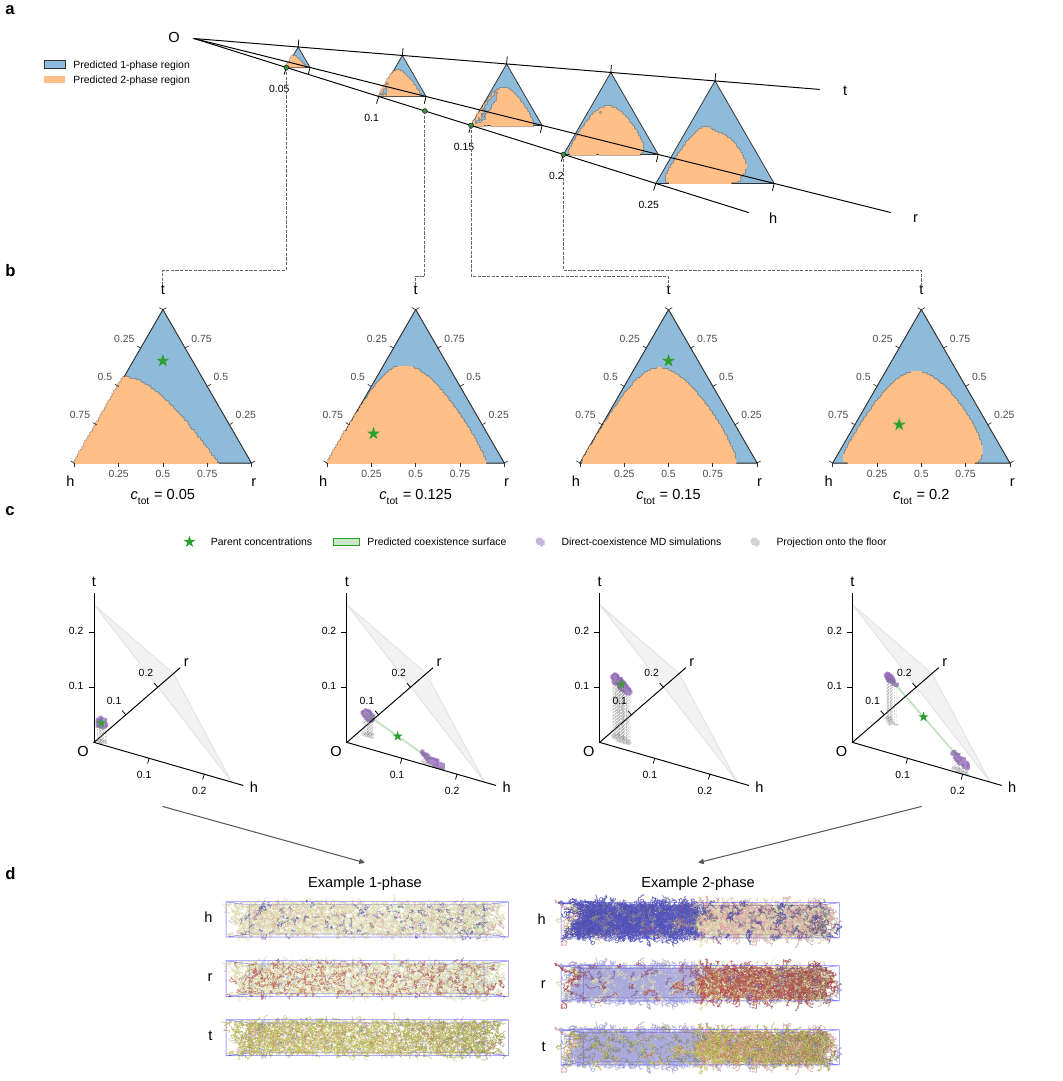}
  \caption[Example ternary phase diagram exhibiting two-phase coexistence.]
          {\textbf{Ternary phase diagram for mixtures of representative sequences {\textsf h}, {\textsf t}, and~{\textsf r}.}
            Panels a--d are as in Fig.~S7, but for a ternary mixture in which the $d=10$ MLP model predicts no three-phase region.
            (b)~Four parent concentration vectors were selected for coexistence simulations.  Two selected mixtures are predicted to form two phases and two are predicted to remain homogeneous.
            (d)~Snapshots from direct-coexistence simulations of selected mixtures.  In each row, one species is shown opaque.}
  \label{fig:S8}
\end{figure}
\clearpage

\begin{figure}[p]
  \centering
  \includegraphics[width=\linewidth]{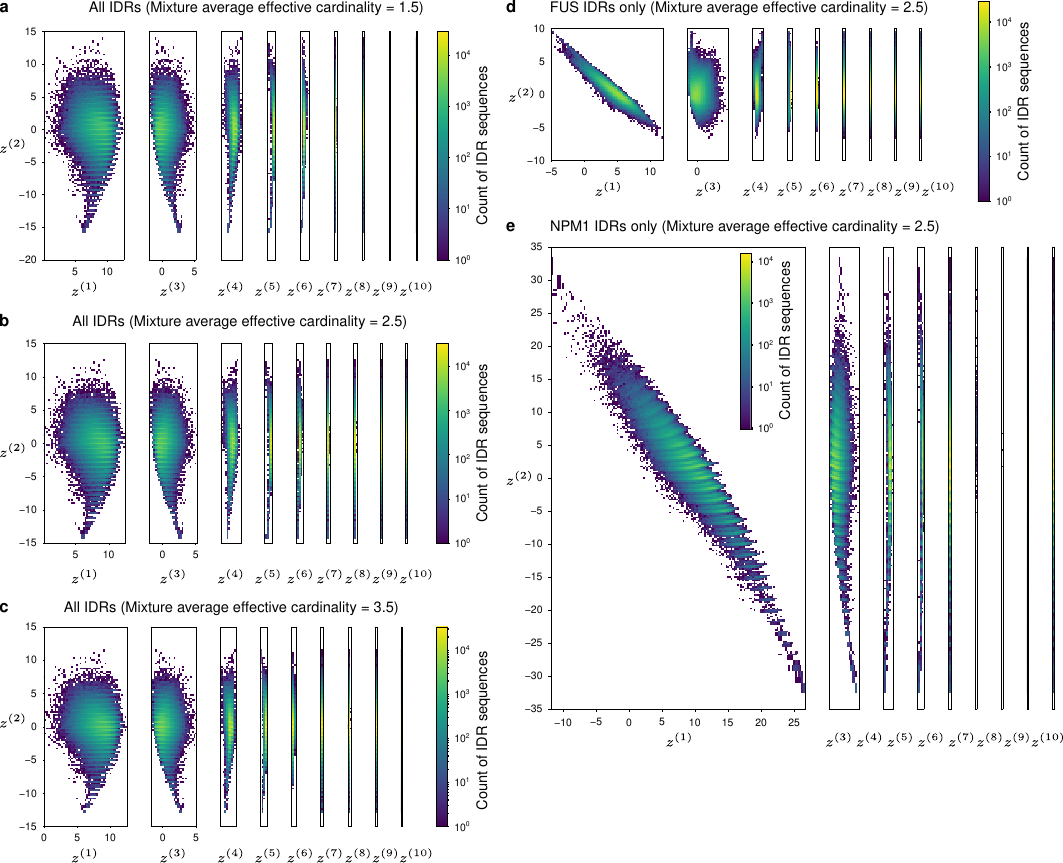}
  \caption[Metric spaces under alternative mixture priors.]
          {\textbf{Distributions of human IDRome sequence fragments under alternative mixture priors.}
            All panels show two-dimensional projections of feature vectors in the $d=10$ MLP-model metric space under the specified mixture prior, with all distances reported in units of $kT$.
            (a--c)~Metric spaces with mixture priors in which IDRs are sampled uniformly from the sequence ensemble.  The Dirichlet distribution sparsity parameter is tuned such that the average effective cardinality ($\sim 99$\% mass) across mixtures is (a)~1.5, (b)~2.5, or (c)~3.5.  The metric space shown in (b) is used throughout the main text and also shown in Fig.~2c.  Increasing the effective cardinality (i.e., reducing mixture sparsity) tends to reduce the spread of the distribution, decreasing thermodynamic distances between IDRs.  However, the qualitative properties of these feature-vector distributions remain consistent, indicating that the choice of the sparsity parameter plays a secondary role relative to the choice of the underlying sequence distribution.
            (d,e)~By contrast, changing the sequence distribution from which the mixture prior is composed significantly alters distances between IDRs.  Metric spaces with mixture priors in which only (d)~FUS fragments or (e)~NPM1 fragments are sampled correspond to the context-dependent mutational analyses shown in Fig.~5j,k.  These mixture priors both have an average effective cardinality of 2.5.
            Importantly, changing the mixture prior involves a linear transformation of the metric space and thus does not require model retraining.
          }
  \label{fig:S9}
\end{figure}
\clearpage

\begin{figure}[p]
  \centering
  \includegraphics[width=\linewidth]{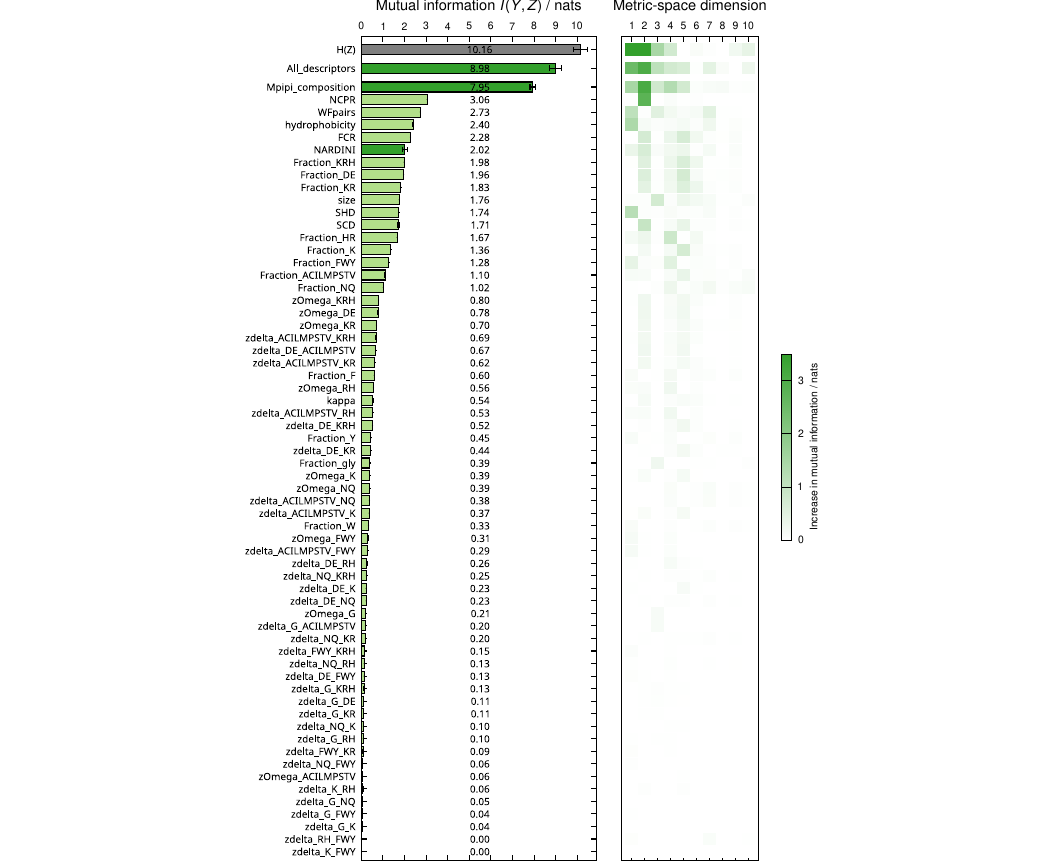}
  \caption[Complete mutual information analysis of theoretical sequence descriptors.]
          {\textbf{Complete mutual information analysis of theoretical sequence descriptors.}
            \textit{Left:}~Mutual information between each individual descriptor (light green) or set of descriptors (dark green) and the $d=10$ MLP-model metric space.  Error bars indicate the standard deviation of five independent estimates.  The entropy of the metric space, $H(Z)$, is shown for comparison (gray).
            \textit{Right:}~For each descriptor or descriptor set, the difference between the mutual information relative to the metric space truncated at dimension $d$ and the mutual information relative to the metric space truncated at dimension $d-1$.}
  \label{fig:S10}
\end{figure}
\clearpage

\begin{figure}[p]
  \centering
  \includegraphics[width=\linewidth]{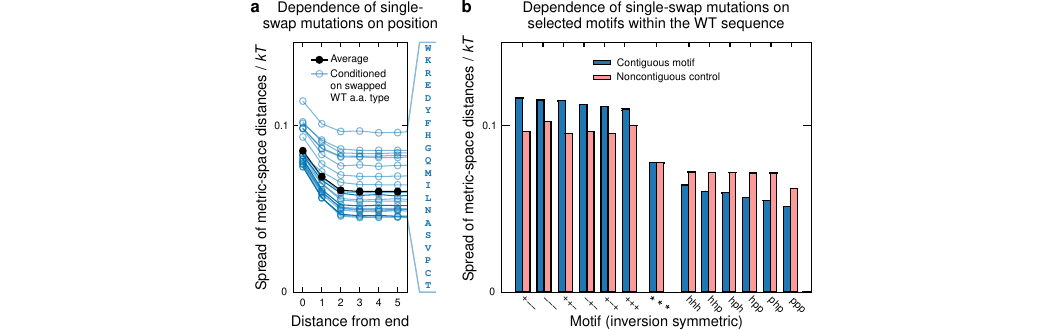}
  \caption[Counterfactual analysis of single-swap mutants without ESM2 weighting.]
          {\textbf{Counterfactual analysis of single-swap mutants without ESM2 weighting.}
            The analyses shown in Fig.~7c,d were repeated without weighting single-swap mutants according to their ESM2 pseudo-likelihood; consequently, all unique single-swap mutant sequences around a WT sequence contribute equally in these calculations.
            (a)~Compared with Fig.~7c, the magnitudes of the spreads increase slightly in the absence of ESM2 weighting; however, the apparent end effects and the ordering of the curves conditioned on the WT amino acids at the swapped positions are unaffected.
            (b)~Compared with Fig.~7d, the qualitative differences between motifs defined based on charge versus hydrophobicity are unchanged.  The qualitative behavior of each contiguous motif class relative to its noncontiguous control is similarly unaffected.  The ordering of the magnitudes of the spreads for charge-based motifs differs slightly between calculations with and without ESM2 weighting.}
  \label{fig:S11}
\end{figure}
\clearpage


\noindent \textbf{Supplementary Information for ``A thermodynamic metric quantitatively predicts disordered protein partitioning and multicomponent phase behavior''}

\tableofcontents
\bigskip

\section{Extended methods}

\subsection{Machine-learning framework for free energy prediction}

In our framework, an encoder maps each sequence to a feature vector, the feature vectors of different sequences in a mixture combine linearly, and a decoder predicts the excess free-energy density of the mixture from the mixture representation.
In this section and in what follows, we use the $\tilde\cdot$ notation to indicate that the learned vectors and functions live in an arbitrary coordinate system, which will be transformed later to construct the thermodynamic metric space.

We map each IDR sequence into a learned latent space $\tilde{z}\in \mathbb{R}^{d}$.
Because Mpipi simulations do not distinguish between the N- and C-termini of polypeptide chains, the latent representation is defined as the average of the raw encoder outputs for the original sequence $s$ and its reverse $s^{\mathrm{rev}}$, 
\begin{equation}
\label{eq:encoder}
    \tilde{z} = \tilde{\phi}(s) = \left[\tilde{\phi}_{\mathrm{raw}}(s)+\tilde{\phi}_{\mathrm{raw}}(s^{\mathrm{rev}})\right]/2.
\end{equation}
For a mixture of IDRs, the mixture’s latent representation is given by the concentration-weighted linear combination of its components,
\begin{equation}
  \tilde{x} = \sum_{i=1}^n c_i\tilde{z}_i,
\end{equation}
where $c_i$ is the number concentration of component $i$ and $n$ is the number of components in the mixture.

A decoder then maps the mixture representation $\tilde{x}$ to the excess free-energy density of the mixture, from which we obtain thermodynamic quantities.
We define the excess free-energy density of the mixture as
\begin{equation}
    \hat f^{\mathrm{ex}}=\tilde{\Psi}(\tilde{x})=\tilde{x}\, \tilde{G}(\tilde{x})\, \tilde{x}^{\mathsf T}. 
\end{equation}
where $\tilde{G}(\tilde{x})\in\mathbb{R}^{d\times d}$ is a symmetric matrix-valued analytic function.
All learnable parameters of the decoder enter through $\tilde{G}(\tilde{x})$.
This construction is invariant to permutation of sequences in a mixture and obeys the ideal gas law in the low-concentration limit.
The excess pressure $\hat P^{\mathrm{ex}}$ and excess chemical potential $\hat \mu^{\mathrm{ex}}$ can be derived by taking derivatives of $\hat f^{\mathrm{ex}}$,
\begin{gather}
    \hat P^{\mathrm{ex}}=\tilde{x}\cdot\nabla_{\tilde{x}}\tilde{\Psi}(\tilde{x})-\tilde{\Psi}(\tilde{x}),\\
    \hat \mu_i^{\mathrm{ex}} = \tilde{z}_i \cdot \nabla_{\tilde{x}}\tilde{\Psi}(\tilde{x}). 
\end{gather}

\subsection{Proof of completeness}

In this section, we show that, for any finite ensemble of molecular sequences, the prescribed encoder–decoder form can represent any analytic excess free-energy density consistent with thermodynamic extensivity, provided that the latent-space dimension $d$ is sufficiently large and the decoder is sufficiently expressive.

Because the excess free energy $F^{\mathrm{ex}}$ of a finite mixture is invariant under permutations of molecular labels, it is a permutation-invariant function of the constituent molecular sequences.
Following the Deep Sets representation of Zaheer et al.,\cite{zaheer2017deep} such a function admits a decomposition of the form
\begin{equation}
\label{eq:zaheer}
F^{\mathrm{ex}}=\tilde{\psi}_V\left(\sum_{j=1}^M\tilde{\phi}_V(s_j)\right),
\end{equation}
where $s_j$ denotes the sequence of the $j$-th molecule, $\tilde z_j=\tilde{\phi}_V(s_j)$ is its latent representation, $M$ is the total number of molecules in the mixture, and $\tilde{\phi}_V$ and $\tilde{\psi}_V$ are suitable nonlinear transformations for a given system volume $V$.
Grouping identical molecular species yields
\begin{equation}
F^{\mathrm{ex}}
=\tilde{\psi}_V\left(\sum_{i=1}^{n} N_i \tilde{\phi}_V(s_i)\right)
=\tilde{\psi}_V\left(\sum_{i=1}^{n} c_i V \tilde{\phi}_V(s_i)\right),
\end{equation}
where $n$ is the number of components in the mixture, $N_i$ is the number of molecules of component $i$, and $c_i=N_i/V$ is the number concentration.

To construct a free-energy model that extrapolates to the thermodynamic limit, we impose extensivity of the excess free energy,
\begin{equation}
    F^{\mathrm{ex}}(V,\{N_i\}) = V f^{\mathrm{ex}}(\{c_i\}).
\end{equation}
Therefore,
\begin{equation}
    f^\mathrm{ex}=\frac{1}{V}F^{\mathrm{ex}}=\frac{1}{V}\tilde{\psi}_V\left(\sum_{i=1}^{n} c_i V \tilde{\phi}_V(s_i)\right). 
\end{equation}
Defining
\begin{equation}
    \tilde{\Psi}_V(x)=\frac{1}{V}\tilde{\psi}_V(Vx),
\end{equation}
we obtain
\begin{equation}
    f^{\mathrm{ex}}= \tilde{\Psi}_V\left(\sum_{i=1}^n c_i \tilde{\phi}_V(s_i)\right).
\end{equation}
Since $f^{\mathrm{ex}}$ is intensive, this expression must be invariant with respect to $V$. For any fixed reference volume $V_0$, 
\begin{equation}
    \tilde{\Psi}_V\left(\sum_{i=1}^n c_i \tilde{\phi}_V(s_i)\right)=\tilde{\Psi}_{V_0}\left(\sum_{i=1}^n c_i \tilde{\phi}_{V_0}(s_i)\right)\quad\text{for all $V$}.
\end{equation}
We therefore define 
\begin{equation}
    f^{\mathrm{ex}}=\tilde{\Psi}\left(\sum_{i=1}^n c_i \tilde{\phi}(s_i)\right)\coloneq\tilde{\Psi}_{V_0}\left(\sum_{i=1}^n c_i \tilde{\phi}_{V_0}(s_i)\right).
\end{equation}

Consistency with the ideal-gas limit requires that $f^{\mathrm{ex}}$ vanishes faster than linearly as $c_{\mathrm{tot}}\to0$ at fixed mixture composition.
Since the decoder can be chosen analytic, this implies that, without loss of generality,
\begin{equation}
\tilde{\Psi}(\tilde{x})=\tilde{x}\,\tilde{G}(\tilde{x})\,\tilde{x}^{\mathsf T},
\end{equation}
where $\tilde{G}(\tilde{x})$ is an analytic, input-dependent matrix that captures higher-order interactions while preserving the correct low-density limit.

We then investigate the number of dimensions required for the latent space $\tilde{z}_i=\tilde{\phi}(s_i)$.
Universal representation of permutation-invariant functions over continuously varying set elements generally requires a latent dimension that scales with the maximum set size when the encoder and decoder are continuous.\cite{wagstaff2019on}
In the above formulation of mixtures of distinct molecules, \eqref{eq:zaheer}, the set size is the number of molecules $M$, which is associated with the system size $V$; in the thermodynamic limit, $V\rightarrow\infty$, the number of molecules diverges.
However, for an ensemble of $S$ distinct sequences, choosing $d=S$ and assigning each sequence a different Cartesian basis vector makes the sum of molecular feature vectors equivalent to the vector of species populations, demonstrating that $d=S$ suffices regardless of $M$.
Therefore, the latent dimension required for completeness is bounded above by the number of unique sequences in the sequence ensemble.

We conclude that, for a sufficiently large latent-space dimension $d$, this framework can represent any analytic and extensive excess free-energy density that is permutation invariant with respect to the molecules in a mixture.
In practice, the latent dimension required to achieve a desired prediction accuracy must be determined empirically, as demonstrated in this work.
Attaining high accuracy with finite neural networks further requires sufficiently expressive encoder and decoder architectures, together with thermodynamic training data that adequately constrain the learned free-energy function.

\subsection{Encoder architecture and implementation}

Each IDR sequence is tokenized into discrete residue tokens and processed by a three-layer Transformer raw encoder with two attention heads, a model dimension of 16, and a feed-forward dimension of 48.
The Transformer outputs are mapped to the latent space via a learnable linear projection.
The encoder output is the average of the raw outputs for the original sequence and its reverse, as in \eqref{eq:encoder}.

\subsection{General decoder architecture and implementation}

To implement a general decoder, we parameterize the symmetric matrix-valued function $\tilde{G}(\tilde{x})$ using an artificial neural network.
The $d(d+1) / 2$ independent entries of the upper triangular part of $\tilde{G}(\tilde{x})$ are parameterized by a multilayer perceptron (MLP) with 2 hidden layers of 128 and 64 neurons, respectively.
To ensure $\tilde{G}(\tilde{x})$ is analytic, the SiLU activation function is used.
The full symmetric matrix is constructed by mirroring the upper triangular entries across the diagonal.

Given $\tilde{G}(\tilde{x})$, the excess free-energy density is computed as $\tilde{\Psi}(\tilde{x})=\tilde{x}\,\tilde{G}(\tilde{x})\,\tilde{x}^{\mathsf T}$, and $\hat{P}^{\mathrm{ex}}$ and $\hat{\mu}^{\mathrm{ex}}$ are calculated by automatic differentiation.
We refer to the model implemented using this general decoder as the ``MLP model'' in the main text.

\subsection{Pairwise decoder architecture and implementation}
\label{sec:pairwise}
The pairwise decoder is a specific, constrained form of the general decoder designed to recover a pairwise interaction structure.
In this formulation, all sequence-dependent contributions to the free energy enter through pairwise interactions between latent sequence representations, analogous to classical mean-field models such as Flory--Huggins (FH).
However, unlike conventional FH parameterizations based on monomer-level interactions, our pairwise coefficients describe effective interactions between entire polymer sequences and can therefore account for sequence-dependent patterning effects.
Higher-order concentration effects are retained only through powers of the total concentration, representing sequence-independent contributions associated with excluded-volume interactions.

The excess free-energy density within this pairwise approximation is
\begin{equation}
\label{eq:fpw}
\hat{f}^{\mathrm{ex}}
=
\frac{1}{2}\tilde{x}\,\tilde{G}^{\mathrm{PW}}\tilde{x}^{\mathsf T}
+\sum_{p=3}^{20} \frac{\tilde{a}_p}{p-1} (\tilde{x}^{(d)})^{\,p}.
\end{equation}
The symmetric matrix $\tilde{G}^{\mathrm{PW}}\in\mathbb{R}^{d\times d}$ is learned and encodes interaction rules between latent dimensions.
In this decoder, $\tilde{G}^{\mathrm{PW}}$ is taken to be constant, independent of $\tilde{x}$.
The final latent dimension $d$ represents a molecular-size coordinate, which is fixed at $\tilde z_i^{(d)}=1$ for all $i$, since all IDR fragments in the sequence ensemble have the same length in this work.
As a result, the second term in \eqref{eq:fpw}, which represents sequence-independent higher-order concentration effects, depends only on the total concentration.
Specifically, $\tilde{x}^{(d)} = \sum_{i=1}^n c_i = c_{\mathrm{tot}}$, with learned coefficients $\tilde{a}_p > 0$.
These choices ensure that all sequence-dependent contributions to $\hat{f}^{\mathrm{ex}}$ enter through the quadratic form $\tilde{x}\,\tilde{G}^{\mathrm{PW}}\tilde{x}^{\mathsf T}$.

For a $d$-dimensional pairwise model, we can write the pairwise interaction coefficient between components $i$ and $j$ in the recognizable form
\begin{equation}
\varepsilon_{ij}=\tilde{z}_i\,\tilde{G}^{\mathrm{PW}}\tilde{z}_j^{\mathsf T},
\end{equation}
so that
\begin{equation}
\frac{1}{2}\tilde{x}\,\tilde{G}^{\mathrm{PW}}\tilde{x}^{\mathsf T}
=
\frac{1}{2}\sum_{i=1}^n \sum_{j=1}^{n}\varepsilon_{ij}c_ic_j.
\end{equation}
From \eqref{eq:fpw}, the excess pressure is obtained as
\begin{equation}
\begin{aligned}
\hat{P}^{\mathrm{ex}}
&=
\tilde{x}\cdot\nabla_{\tilde{x}}\hat{f}^{\mathrm{ex}}
-\hat{f}^{\mathrm{ex}} \\
&=
\frac{1}{2}\sum_{i=1}^n \sum_{j=1}^{n}\varepsilon_{ij}c_ic_j
+\sum_{p=3}^{20} \tilde{a}_p (\tilde{x}^{(d)})^{\,p}.
\end{aligned}
\end{equation}
Including the ideal contribution $P^{\mathrm{id}}=c_{\mathrm{tot}}kT$, the total pressure takes the form
\begin{equation}
\hat{P}
=
c_{\mathrm{tot}}kT
+
\frac{1}{2}\sum_{i,j=1}^{n}\varepsilon_{ij}c_ic_j
+\sum_{p=3}^{20} \tilde{a}_p (\tilde{x}^{(d)})^{\,p},
\end{equation}
which matches the structure of FH-type models: an ideal term, a quadratic pairwise interaction term, and additional higher-order concentration-dependent corrections.

Accordingly, the excess chemical potential of component $i$ is obtained from \eqref{eq:fpw} as
\begin{equation}
\begin{aligned}
\hat{\mu}_i^{\mathrm{ex}}
&=
\frac{\partial \hat{f}^{\mathrm{ex}}}{\partial c_i} \\
&=
\sum_{j=1}^{n}\varepsilon_{ij}c_j
+\sum_{p=3}^{20} \frac{p}{p-1}\tilde{a}_p (\tilde{x}^{(d)})^{\,p-1}.
\end{aligned}
\end{equation}
Thus, differences between the excess chemical potentials of distinct components depend only on the pairwise interaction coefficients, since the higher-order concentration-dependent contributions are identical for all species within the pairwise model.

\subsection{Latent space transformation and alignment}
\label{sec:latent_space_transformation}

This section describes a physically motivated linear transformation of the latent space of each trained model, defined by the $L^2(\rho_{\mathrm{mix}})$ norm of the predicted excess chemical potential $\hat \mu^{\mathrm{ex}}$.
This norm endows the latent space $\tilde z \in \mathbb{R}^d$ of each trained model with a consistent metric, producing a shared metric space $z \in \mathbb{R}^d$ in which the positions of sequences correspond directly to the physical $\hat\mu^{\mathrm{ex}}$ geometry.
We first introduce a principled strategy for sampling mixtures.
We then define a metric space such that Euclidean distances equal the norms of differences between $\hat\mu^{\mathrm{ex}}$ functions.
Lastly, we canonicalize the metric-normalized coordinates using orthogonal transformations around the origin, resulting in a single aligned metric space across all trained models.

\subsubsection{Mixture prior}

We consider mixtures of $n$ different sequences sampled from the IDR fragment ensemble with concentrations $c_i \ge 0$ and total concentration $c_{\mathrm{tot}} = \sum_{i=1}^n c_i$.
We parameterize mixtures by the total concentration $c_{\mathrm{tot}}$ and mole fractions $\{y_i = c_i / c_{\mathrm{tot}}\}$.
We define the prior over mixtures by independently sampling the total concentration and a mole-fraction vector over all $S$ sequences in the IDR fragment ensemble.
The prior factorizes as $\rho_{\mathrm{mix}}(c_{\mathrm{tot}},y)=p(c_{\mathrm{tot}})p(y)$, where $c_{\mathrm{tot}}\sim\mathrm{Uniform}(0,c_{\mathrm{tot},\max}]$, with $c_{\mathrm{tot}, \max}=\SI{0.25}{\chain\per\nm\cubed}$, and $y\sim\mathrm{Dirichlet}(\tilde\alpha/S,\ldots,\tilde\alpha/S)$.
Here $\tilde\alpha=O(1)$ is a dimensionless total Dirichlet concentration parameter; in the large-$S$ limit this yields sparse mole fractions with an $O(1)$ effective cardinality.
In this work, we chose the value of $\tilde\alpha$ empirically to achieve a desired effective cardinality threshold of $|\{ i : y_i > \varepsilon \} | \sim 2.5$ with $\varepsilon \approx 10^{-2}$.
This prior distribution places support over all mixtures that can be composed from the $S$ sequences in the IDR fragment ensemble, while biasing toward relatively sparse mole fractions in which only a small subset of sequences is strongly enriched.

To sample mole fractions, we use a stick-breaking (GEM) construction\cite{sethuraman1994constructive} with parameter $\tilde\alpha$:
\begin{equation}
v_i \sim \mathrm{Beta}(1, \tilde\alpha),
\qquad
w_i = v_i \prod_{j<i} (1 - v_j),
\end{equation}
where the weights $w_i$ represent the mole fractions of the selected sequences. The construction is truncated once the remaining mass is below a tolerance of $10^{-6}$.
The weights $\{w_i\}$ are renormalized and assigned to randomly chosen distinct sequence indices.
This procedure approximates the large-$S$ limit of a symmetric Dirichlet distribution with concentration parameters $\tilde\alpha/S$.

\subsubsection{Metric definition}

For each trained model $k = 1, \dots, K$, we define
\begin{equation}
  M^{(k)}=\mathbb E_{c\sim\rho_{\mathrm{mix}}}\left[\nabla\tilde\Psi^{(k)}(\tilde x^{(k)}(c))\nabla\tilde\Psi^{(k)}(\tilde x^{(k)}(c))^{\mathsf T}\right],
\end{equation}
which is symmetric positive semidefinite and induces the $L^2(\rho_{\mathrm{mix}})$ norm
\begin{equation}
\|\hat \mu_i^{\mathrm{ex},(k)}\|_{L^2(\rho_{\mathrm{mix}})}^2
=
\tilde{z}_i^{(k)} M^{(k)} (\tilde{z}_i^{(k)})^{\mathsf T}.
\end{equation}
We estimate $M^{(k)}$ by Monte Carlo sampling from the prior, $\rho_{\mathrm{mix}}$.
For each draw, we sample $y$ by stick breaking, sample $c_{\mathrm{tot}}$, form 
\begin{equation}
\tilde{x}^{(k)} = c_{\mathrm{tot}} \sum_{i=1}^n (c_i / c_{\mathrm{tot}}) \tilde{z}_i^{(k)},
\end{equation}
evaluate $g^{(k)}=\nabla_{\tilde x}\tilde\Psi^{(k)}(\tilde x^{(k)})$, and accumulate
$g^{(k)} (g^{(k)})^{\mathsf T}$.
Averaging over samples yields the Monte Carlo estimate $\widehat M^{(k)}$.

The exact metric transformation is defined by the principal square root
\begin{equation}
  A^{\mathrm{trans},(k)}=(M^{(k)})^{1/2},
\end{equation}
yielding transformed latent coordinates
\begin{equation}
z_i^{\mathrm{trans}, (k)}
=
\tilde{z}_i^{(k)} (A^{\mathrm{trans}, (k)})^{\mathsf T}
\end{equation}
that satisfy
\begin{equation}
\|z_i^{\mathrm{trans}, (k)}\|_2^2
=
\tilde{z}_i^{(k)} M^{(k)} (\tilde{z}_i^{(k)})^{\mathsf T}.
\end{equation}
We assume that the metric matrix is positive definite on the retained latent dimensions, so that the transformation is invertible.
In numerical calculations, we approximate this transformation using the Monte Carlo estimate $(\widehat M^{(k)})^{1/2}$.

\subsubsection{Canonicalization and alignment}

For each model, we perform uncentered principal component analysis (PCA) on ${z}^{\mathrm{trans}, (k)}$.
Specifically, we diagonalize $(1/S) (Z^{\mathrm{trans}, (k)})^{\mathsf T} Z^{\mathrm{trans}, (k)}$, where the rows of $Z^{\mathrm{trans}, (k)}$ are $z_i^{\mathrm{trans}, (k)}$ for all sequences $i$ in the IDR fragment ensemble, rotate into the PCA basis, and fix the sign of each axis by requiring the mean coordinate to be nonnegative.
This removes arbitrary rotations and reflections within each model.

Starting from the PCA-canonicalized coordinates, we note that for a perfect model the resulting metric space is already unique up to an orthogonal transformation (see \secref{sec:uniqueness}).
Thus, no further alignment should be required in principle. 
In practice, however, small discrepancies arise because the trained models are imperfect.
To reduce residual misalignment across the $K$ models, we apply generalized orthogonal Procrustes to optimize their alignment in a common frame.
We find that this step results in only minor rotational adjustments, since the per-model PCA-canonicalized coordinates are already extremely similar.

We then define a global canonical basis by applying uncentered PCA to the mean (aligned) model, with the same sign convention, and rotate all models into this basis.
The resulting aligned coordinates satisfy
\begin{equation}
z_i^{(k)}
=
\tilde{z}_i^{(k)} (A^{(k)})^{\mathsf T},
\end{equation}
where $A^{(k)}=O^{(k)}A^{\mathrm{trans}, (k)}$ folds together the metric square root and all subsequent orthogonal transformations. For each model, the excess free-energy density in the aligned coordinate system becomes
\begin{equation}
    \hat{f}^{\mathrm{ex}, (k)} = \Psi^{(k)}(x^{(k)}),
\end{equation}
where
\begin{gather}
    x^{(k)}=\sum_ic_iz_i^{(k)}=\tilde{x}^{(k)} (A^{(k)})^{\mathsf T}, \\
\Psi^{(k)}(x^{(k)})
=
x^{(k)}(A^{(k)})^{-\mathsf T}\,
\tilde{G}^{(k)}\!\left(x^{(k)}(A^{(k)})^{-\mathsf T}\right)\,
(A^{(k)})^{-1}\,(x^{(k)})^{\mathsf T}.
\end{gather}
The encoder in the transformed coordinate system becomes
\begin{equation}
    \phi^{(k)}(s_i)=z^{(k)}_i=\tilde{\phi}^{(k)}(s_i)(A^{(k)})^\mathsf T.
\end{equation}
For perfect models, the canonicalized feature vectors should agree up to any residual orthogonal symmetries, with equivalent free-energy predictions over the physically accessible mixture domain (see \secref{sec:uniqueness}).

We refer to the mean coordinates in this aligned \textit{thermodynamic metric space} obtained from the trained general-decoder MLP models as 
\begin{equation}
  z_i = \frac{1}{K} \sum_{k=1}^K z_i^{(k)}
\end{equation}
in the main text, where we average over the $K=5$ models from five-fold cross-validation, and we report the uncertainty of each point in this metric space as
\begin{equation}
  \label{eq:RMSD_i}
  \textrm{RMSD}_i = \sqrt{\frac{1}{K} \sum_{k=1}^K \left\|z_i^{(k)} - z_i \right\|_2^2}.
\end{equation}
We obtain ensemble-averaged excess free-energy predictions by evaluating each decoder using its corresponding transformed mixture representation and averaging the resulting predictions.

Because the $K=5$ models were obtained by five-fold cross-validation, their training datasets overlap.
Consequently, the RMSD defined in \eqref{eq:RMSD_i} quantifies the observed variability of feature vectors across independently optimized encoders trained on overlapping datasets, rather than providing a calibrated confidence interval or a comprehensive estimate of model uncertainty.
In particular, this measure does not necessarily capture uncertainty associated with independent resampling of training data or systematic errors in the learned model.
Comparisons between counterfactual mutation spreads and the cross-validation RMSD, shown in Fig.~7a in the main text, should therefore be interpreted as relative measures of variability within the trained model ensemble, rather than formal statistical significance tests.

\subsection{Proof of uniqueness}
\label{sec:uniqueness}

For any pair of points $i,j$ in the $z$-space of model $k$, we have
\begin{equation}
\label{eq:distance_equivalence}
\begin{aligned}
\|{z}_i^{(k)}-{z}_j^{(k)}\|_2
&=
\sqrt{(\tilde{z}_i^{(k)}-\tilde{z}_j^{(k)})(A^{\mathrm{trans},(k)})^{\mathsf T} (O^{(k)})^{\mathsf T}O^{(k)}A^{\mathrm{trans},(k)}(\tilde{z}_i^{(k)}-\tilde{z}_j^{(k)})^{\mathsf T}}
\\
&=
\sqrt{(\tilde{z}_i^{(k)}-\tilde{z}_j^{(k)}) M^{(k)} (\tilde{z}_i^{(k)}-\tilde{z}_j^{(k)})^{\mathsf T}}
\\
&=
\|\hat{\mu}_i^{\mathrm{ex},(k)}-\hat{\mu}_j^{\mathrm{ex},(k)}\|_{L^2(\rho_{\mathrm{mix}})} .
\end{aligned}
\end{equation}
When the model accurately predicts excess chemical potentials, the right-hand side of \eqref{eq:distance_equivalence} depends only on the physical quantities and is therefore invariant across different models $k = 1, \ldots, K$.
Consequently, pairwise Euclidean distances between points in the corresponding $z$-spaces are identical for all exact models.

In addition, $z_i^{(k)} = \vec{0}$ indicates $\|\mu_i^{\mathrm{ex},(k)}\|_{L^2(\rho_{\mathrm{mix}})} = 0$, which is true if and only if $\mu_i^{\mathrm{ex},(k)}=0$ for all mixtures.
We therefore identify the origin in $z$-space with an ideal, noninteracting reference polymer, whose excess chemical potential vanishes across the mixture environments described by the prior.

These results imply that the embeddings produced by different accurate models preserve both pairwise distances and distances from the origin.
Any two such embeddings are therefore related by a distance-preserving transformation that fixes the origin.
These transformations form the orthogonal group $O(d)$, consisting of rotations and reflections.

Hence, the thermodynamic metric representation is unique up to an orthogonal transformation.
The quantity denoted as $z$ in the main text is the ensemble average of approximately aligned representations, whose coordinate orientations are fixed using uncentered PCA.

\subsection{Curation of IDR fragments}
\label{sec:idr_curation}

IDR fragments were collected from a dataset of 28,058 full-length IDRs spanning 15,347 distinct proteins in the human proteome, which was previously reported by Tesei et al.\cite{tesei2024conformational}
For each full-length IDR, fragment extraction was initiated with a 20-residue segment (20-mer) centered within the sequence.
Additional 20-mers were subsequently generated by translating the 20-residue window in both the N-terminal and C-terminal directions using a step size of 10 residues, continuing until insufficient residues remained to construct a new 20-mer.
This procedure avoids extracting neighboring 20-mers with greater than 50\% positional overlap in any single full-length IDR sequence by construction.
In this way, we generated a total of 335,439 20-mers, which were then randomly divided into a training set of 235,439 20-mers for model training and validation and a test set of 100,000 20-mers.
We verified that our results are insensitive to alternative train/test splits of the IDR fragment ensemble (see \secref{sec:splits}).

\subsection{Generation of mutant fragments}
\label{sec:sequence_generation}

We generated mutant sequences for each wild-type (WT) sequence in the test set.
Whereas the WT IDR fragments are empirical samples from an underlying sequence distribution $\rho_{\mathrm{seq}}$, uniform sampling of mutations does not in general produce mutant (MUT) sequences from the same distribution.
We therefore estimated $\ln\left(
{\rho_{\mathrm{seq}}(s_{\mathrm{MUT}})}
     /{\rho_{\mathrm{seq}}(s_{\mathrm{WT}})}
\right)$ by evaluating the difference between the pseudo log-likelihoods (PLL) of the MUT and WT sequences in the protein language model ESM2 (esm2\_t6\_8M\_UR50D).\cite{lin2023evolutionary}
To avoid ESM2 artifacts involving the ends of sequences, we performed PLL calculations with sequences embedded in their native context within the full-length human IDR.
We used a fixed context window extending three positions at either end of the sequence for this purpose.
For sequences with insufficient native context, we filled the context window by padding with polyglycine.
We then either sampled multiple-mutation sequences using a Markov chain Monte Carlo (MCMC) scheme in accordance with the ESM2 proxy for $\rho_{\mathrm{seq}}$, or we exhaustively enumerated all point-mutation or unique single-swap mutant sequences and used the ESM2 proxy for $\rho_{\mathrm{seq}}$ to weight each sequence in the calculation of sequence-ensemble averages, as described below.

\subsubsection{MCMC sequence generation of multi-swap mutants}

We used MCMC to stochastically sample composition-preserving variants of each WT IDR fragment in the test set.
This scheme is designed to explore the space of local rearrangements of the WT residues while remaining close to the WT sequence and avoiding sequences that are implausible under the language model.

\paragraph{Loss function.}
Let $\mathcal{A}$ denote the standard 20-amino-acid alphabet, and let
\begin{equation}
s = (s_1,\dots,s_L) \in \mathcal{A}^L
\end{equation}
be a protein sequence of fixed length $L$.
For a given sequence $s$, we compute
\begin{equation}
\mathrm{PLL}(s)
= \sum_{l=1}^L \log p_\theta(s_l \mid s_{-l}, \text{context}),
\end{equation}
where $p_\theta(\cdot)$ is the masked-language model and the sum only runs over positions within the sequence.
Each term in the PLL calculation is conditioned on the sequence at all positions other than the masked position $l$, $s_{-l}$, as well as the fixed context window extending three positions beyond each end of the sequence.
In addition, we measure similarity to the WT sequence $s_{\mathrm{WT}}$ using the Hamming distance
\begin{equation}
d_{\mathrm{H}}(s, s_{\mathrm{WT}}) = \sum_{l=1}^L \mathbf{1}[s_l \neq (s_{\mathrm{WT}})_l].
\end{equation}
We then define the hinge loss
\begin{equation}
H(s) = \max\!\left(0,\, \mathrm{PLL}(s_{\mathrm{WT}}) - \mathrm{PLL}(s)\right),
\end{equation}
which penalizes sequences whose PLL falls below that of the WT, but does not reward sequences with higher-than-WT PLL.
This prevents the chain from drifting toward overly predictable model attractors.
The MCMC sampler targets the unnormalized distribution
\begin{equation}
\pi(s) \propto
\exp\!\left(
-\frac{H(s) + \lambda\, d_H(s, s_{\mathrm{WT}})}{T_{\mathrm{MCMC}}}
\right),
\end{equation}
where $T_{\mathrm{MCMC}}>0$ is a fictitious temperature parameter and $\lambda \ge 0$ controls the strength of the penalty for deviating from the WT.

\paragraph{Metropolis--Hastings algorithm.}
The state space for MCMC sampling consists of all permutations of the WT sequence residues.
A proposal is generated by sampling $n_{\mathrm{swap}} \sim \mathrm{Uniform}\{1,\dots,4\}$ and applying $n_{\mathrm{swap}}$ random swap moves, each of which exchanges two positions in the sequence.
This proposal preserves amino-acid composition exactly and is symmetric.
Given a current state $s$ and a proposal $s'$, the Metropolis--Hastings acceptance probability is
\begin{equation}
\alpha = \min\!\left[
1,\;
\exp\!\left(
-\frac{[H(s') - H(s)] + \lambda [d_H(s', s_{\mathrm{WT}}) - d_H(s, s_{\mathrm{WT}})]}{T_{\mathrm{MCMC}}}
\right)
\right].
\end{equation}

In the production run, we initialized each chain at the WT sequence, ran for a burn-in period of 200 MCMC steps, and then advanced the chain in fixed-length blocks of 80 MCMC steps to obtain uncorrelated samples.
After each block, the current sequence was recorded if it was distinct from the WT and from previously recorded samples.
Sampling was performed using $T_{\mathrm{MCMC}} = 1$ and $\lambda = 1.75$ and continued until nine unique MUT sequences were generated per WT sequence.

\subsubsection{Unique single-swap generation}

We deterministically enumerated all composition-preserving variants that can be obtained from each WT IDR fragment in the test set via a single swap of two residues.
For a WT sequence $s_{\mathrm{WT}}$, unique single-swap mutants are defined as
\begin{equation}
s_{(l,l')} = \text{swap}(s_{\mathrm{WT}}, l, l'),
\quad \text{for } l<l' \text{ and } (s_{\mathrm{WT}})_l \neq (s_{\mathrm{WT}})_{l'}.
\end{equation}
Each resulting sequence differs from the WT by exactly one swap, preserves amino-acid composition, and is guaranteed to be distinct from the WT.
For each mutant, we computed PLL using the same definition as in the MCMC framework.

\subsubsection{Point mutation generation}
We deterministically enumerated all single-residue substitutions (i.e., point mutations) of each WT IDR fragment in the test set.
Unlike the swap-based methods, this procedure does not preserve amino-acid composition, but instead provides a complete local landscape of single-site effects.
For each position $l \in \{1,\dots,L\}$ and each amino acid
$a \in \mathcal{A} \setminus \{(s_{\mathrm{WT}})_l\}$,
a mutant $s_{(l \to a)}$ is generated by substituting the residue at position $l$ with $a$.
This procedure yields exactly $19L$ mutants per WT IDR fragment.
For each mutant, we computed PLL using the same definition as in the MCMC framework.

\subsection{Coarse-grained force fields}

\subsubsection{Mpipi force field}
\label{sec:mpipiFF}

Equation-of-state, free-energy, and phase-coexistence simulations were conducted using the Mpipi force field\cite{joseph2021physics} implemented in LAMMPS.\cite{thompson2022lammps}
Mpipi is a residue-level coarse-grained model for predicting biomolecular phase behavior.
In the Mpipi model, each amino acid is represented by a single bead, with specified mass, diameter ($\sigma$), charge ($q$), and an energy scale ($\varepsilon$) for short-range interactions.\cite{joseph2021physics}
The potential energy is specified as
\begin{equation}
\label{eq:Mpipi_potential}
E_{\mathrm{Mpipi}} = E_{\mathrm{bond}} + E_{\mathrm{elec}} + E_{\mathrm{pair}} .
\end{equation}
The bond energy is given by a harmonic bond potential 
\begin{equation}
\label{eq:Mpipi_potential_bond}
E_{\mathrm{bond}} = \sum_{\mathrm{bonds}\;i} \frac{1}{2}k(r_i-r_{i,\mathrm{ref}})^2,
\end{equation}
where $k$ is set to \SI{8.03}{\joule\per\mole\per\pm\squared}, $r_i$ is the bond length, and the reference bond length $r_{i,\mathrm{ref}}$ is \SI{381}{\pm}.
The electrostatic interactions are modeled using a Coulomb potential with Debye--H\"uckel screening
\begin{equation}
\label{eq:Mpipi_potential_electrostatic}
E_{\mathrm{elec}} = \sum_{i,j} \frac{q_iq_j}{4\pi\varepsilon_r\varepsilon_0r_{ij}}\exp(-\kappa r_{ij}),
\end{equation}
where $\varepsilon_r=80$ is the relative dielectric constant of water, $\varepsilon_0$ is the vacuum permittivity, and $\kappa^{-1}=\SI{795}{\pm}$ is the Debye screening length corresponding to a monovalent salt concentration of \SI{0.15}{\mole\per\liter}.
The cutoff distance of $E_{\mathrm{elec}}$ is set to \SI{3.5}{\nm}.

Nonbonded interactions between amino-acid beads are described by the Wang--Frenkel (WF) potential, which for a pair of beads of types $a$ and $b$ separated by a distance $r$, is given by 
\begin{equation}
\label{eq:Mpipi_potential_WF}
\phi_{ab}(r) = \varepsilon_{ab} \left [ \left ( \frac{\sigma_{ab}}{r} \right )^{2\mu_{ab}}-1 \right ] \left [ \left ( \frac{R_{ab}}{r} \right )^{2\mu_{ab}}-1\right ]^{2\nu_{ab}}\!\!,
\end{equation}
with $\sigma_{ab}$, $\varepsilon_{ab}$, and $\mu_{ab}$ uniquely defined for each pair of interacting bead types.
The parameters $\nu_{ab}$ and $R_{ab}$ are set to 1 and 3$\sigma_{ab}$, respectively.
The total pairwise nonbonded potential energy $E_{\mathrm{pair}}$ is computed by summing $\phi_{ab}$ for all pairs of beads within the corresponding cutoff distances $R_{ab}$; nonbonded interactions between directly bonded residues are excluded.

\subsubsection{CALVADOS force field}
\label{sec:calvadosFF}

Equation-of-state simulations of the WT and MUT binary-mixture test sets were also performed using the CALVADOS force field\cite{tesei2021accurate} implemented in OpenMM\cite{eastman2024openmm8} (Figs.~3c,d and Fig.~S5).
CALVADOS is an alternative residue-level coarse-grained model in which the potential energy is also decomposed into bonded, electrostatic, and pairwise nonbonded contributions, as in \eqref{eq:Mpipi_potential}.
The bonded interactions are described by the harmonic potential in \eqref{eq:Mpipi_potential_bond}, with
$k=\SI{8.03}{\joule\per\mole\per\pm\squared}$ and reference bond lengths determined by the residue-specific parameters of CALVADOS.
Electrostatic interactions are described by \eqref{eq:Mpipi_potential_electrostatic}.
Simulations were performed at $T=\SI{300}{\kelvin}$ and an ionic strength of \SI{0.1464}{\mole\per\liter}, with a relative dielectric constant $\varepsilon_r=77.7$ and an electrostatic cutoff set to \SI{2.5}{\nm}.
Unlike in Mpipi simulations, charges are assigned to both chain termini.

Short-range interactions in the CALVADOS model are described by the Ashbaugh--Hatch (AH) potential,
\begin{equation}
\label{eq:CALVADOS_potential_AH}
\phi_{ab}(r) = \begin{cases}
  \phi_{\mathrm{LJ}}(r) - \lambda_{ab}\phi_{\mathrm{LJ}}(r_{\mathrm{c}}) + (1-\lambda_{ab})\varepsilon,
  & r \leq 2^{1/6}\sigma_{ab}, \\[4pt]
  \lambda_{ab}
  \left[ \phi_{\mathrm{LJ}}(r) - \phi_{\mathrm{LJ}}(r_{\mathrm{c}}) \right],
  & 2^{1/6}\sigma_{ab}<r<r_{\mathrm{c}}, \\[4pt]
  0,
  & r\geq r_{\mathrm{c}},
\end{cases}
\end{equation}
where
\begin{equation}
\label{eq:CALVADOS_potential_LJ}
\phi_{\mathrm{LJ}}(r) = 4\varepsilon \left[ \left(\frac{\sigma_{ab}}{r}\right)^{12} - \left(\frac{\sigma_{ab}}{r}\right)^6 \right].
\end{equation}
Here, $\sigma_{ab}=(\sigma_a+\sigma_b)/2$ and $\lambda_{ab}=(\lambda_a+\lambda_b)/2$ are arithmetic averages of residue-specific diameters and hydropathy parameters, respectively.
The interaction energy scale is $\varepsilon=\SI{0.2}{\kcal\per\mole}$, and the AH cutoff is set to \SI{2.0}{\nm}.
Nonbonded interactions between directly bonded residues are excluded.

\subsection{Equation-of-state simulations}

Equation-of-state (EOS) simulations were performed in the canonical ensemble for $n$-component IDR systems with $n \in \{1, 2, 3, 4\}$.
In each simulation, 100 IDR molecules were initialized uniformly in a cubic simulation box at the prescribed mole fractions and total concentration.
After energy minimization, the system was simulated for \SI{10}{\ns} with a timestep of \SI{10}{\fs} using Velocity Verlet integration.
Pressure was recorded every 1,000 timesteps and averaged after a typical equilibration time of \SI{2}{\ns} to generate the EOS data.

All simulations were carried out at \SI{300}{\kelvin} using a Langevin thermostat with a damping parameter of \SI{1}{\ps} under periodic boundary conditions.
The temperature, timestep, and thermostat scheme were chosen to be consistent with prior IDR simulations reported by Joseph et al.\cite{joseph2021physics}

\subsection{Model training, validation, and EOS testing procedures}
\label{sec:train_val_test}

To generate the training and validation datasets, we performed simulations for 20,000 unary and 120,000 binary IDR systems (see \secref{sec:resource}).
IDR sequences were randomly drawn from the curated library of 20-mer fragments.
For the 20,000 unary systems and 100,000 of the binary systems, the total concentration was randomly sampled from a piecewise-uniform distribution with 20\% probability in the range \SIrange{1e-5}{1e-3}{\chain\per\nm\cubed}, 20\% in \SIrange{1e-3}{0.01}{\chain\per\nm\cubed} and 60\% in \SIrange{0.01}{0.25}{\chain\per\nm\cubed}.
Total concentrations in the other 20,000 binary systems were sampled exclusively from the range \SIrange{0.01}{0.25}{\chain\per\nm\cubed}.
For binary mixtures, this total concentration was randomly partitioned between the two sequence types, resulting in uniformly sampled mole fractions.

Models were trained using five-fold cross-validation.
The simulated dataset was randomly partitioned into five folds, with four folds used for training and the remaining fold used for validation (see \secref{sec:splits} for a discussion of alternative train/test splits).
This procedure yielded five models, each validated on a different fold, for each latent-space dimension $d$.
For each split, we trained five replicate models with different random initializations (random seeds) and retained the replicate with the lowest validation loss.
Models were trained in PyTorch\cite{paszke2019pytorch} and optimized using the Adam optimizer.\cite{zhang2018improved} The transformer encoder used dropout = 0.1.
We set the batch size to 2,048/4,096 and tuned the learning rate using a grid search over \{0.001, 0.005, 0.01\}/\{0.0001, 0.001, 0.01, 0.1\} for the MLP/PW models.
For the PW model, two separate optimizers were used for updating the parameters of $\tilde{G}^\mathrm{PW}$ and the $\tilde{a}_p$, respectively (see \secref{sec:pairwise}).
We used MSE($P/c_\mathrm{tot}$) as the training loss to emphasize accuracy at low total concentration, which is important for reliable free-energy prediction; moreover, with this choice, the simulation uncertainties are approximately independent of the total concentration (Fig.~S1a--c).
Training used early stopping based on validation loss, and the checkpoint with the lowest validation loss was retained.

For testing, we constructed two classes of datasets in which test-set sequences were drawn either uniformly from the sequence pool or uniformly from the metric space.
For each class, we generated test sets for single-component, two-component, three-component, and four-component mixtures.
Test-set total concentrations were sampled as in the 20,000 unary and 100,000 binary systems in the training set.
Random mole fractions were then sampled by randomly partitioning the total concentration among the $n$ sequence types.

\subsection{Calculation of free-energy differences from simulation} 

Free-energy test data for selected $n$-component IDR systems, with $n \in \{1, 2, 3, 4\}$, were obtained by integrating $P^{\mathrm{ex}}$ simulation data.
For each $n$-component IDR system, two mole-fraction vectors were randomly selected, except for single-component systems, in which case the single component's mole fraction is always one.
EOS simulations were performed for each of the mole-fraction choices at 80 total concentrations, logarithmically spaced in the range \SIrange{1e-5}{0.25}{\chain\per\nm\cubed}.

For each mole-fraction choice, the ideal contribution ($c_{\mathrm{tot}}kT$) was subtracted from the simulated EOS data to generate the $P^{\mathrm{ex}}(c_{\mathrm{tot}})$ simulation data, which was subsequently fitted with a 100th-order polynomial in $c_{\mathrm{tot}}$ starting from the quadratic term.
The fitted $P^{\mathrm{ex}}(c_{\mathrm{tot}})$ polynomial was then integrated to obtain the excess free-energy density $f^{\mathrm{ex}}$.
The total free-energy density is thus
\begin{equation}
f
=
\underbrace{
kT\sum_{i=1}^n c_i(\ln(c_i)-1)
}_{f^{\mathrm{id}}}
+
\underbrace{
c_{\mathrm{tot}}\int_0^{c_{\mathrm{tot}}}\frac{P^{\mathrm{ex}}}{r^2}\,dr
\vphantom{\sum_{i=1}^n c_i(\ln(c_i)-1)}
}_{f^{\mathrm{ex}}}.
\label{eq:fdecomposition}
\end{equation}

For each IDR system with $n>1$, we applied this procedure to obtain the free-energy density function $f(c_\mathrm{tot})$ along each selected concentration ray of constant mole fractions.
 We then randomly selected one total concentration uniformly from the concentration range along each ray and used the corresponding free-energy densities $f_1$ and $f_2$ to define the free-energy density difference $\Delta f=|f_1-f_2|$.
For each single-component IDR system, only one free-energy density was sampled, and the free-energy density difference was defined as $\Delta f=|f_1-f^{\mathrm{id}}|$.
The simulated $\Delta f$ data formed the test dataset, which was used to evaluate the $\mathrm{RMSE}(\Delta f)$ of the model-predicted free-energy differences, $\widehat{\Delta f}$.

Model performance was quantified by repeating this sampling and evaluation procedure 100 times.
The $\mathrm{RMSE}(\Delta f)$ values and the associated error bars reported in Fig.~2k represent the mean and standard deviation across these realizations.
In the scatter plots shown in Fig.~2j and Fig.~S3d, we used the realization for which the computed learned-model performance is closest to the reported average.

\subsection{Evaluation of alternative train/test splits}
\label{sec:splits}

To verify that our models' predictive performance is not inflated by potential data leakage arising from contiguous residue overlaps between fragments in training and test mixtures (owing to the 10-residue sliding window step size used during fragment extraction), we evaluated our models' test performance using two alternative data partitioning strategies.

We first performed a strict protein-level data split to ensure that no IDR fragments from the same protein appear in both the training set and the test set.
Here, we randomly set aside 2,971 out of the 15,347 human proteins considered in our study exclusively for testing.
We discarded all samples from the original 140,000-element training dataset that contain fragments from these 2,971 test proteins, resulting in a 104,668-element filtered training dataset.
We then retrained the MLP and PW models from scratch using this filtered dataset.

Because the original test datasets contained very few samples consisting exclusively of sequences from the 2,971 test proteins, we generated 1,000 new EOS test simulations for 1-, 2-, 3-, and 4-species systems.
These test systems comprise mixtures of IDR fragments drawn exclusively from the 2,971 test proteins.
The test performances of the retrained models are nearly indistinguishable from those of the original models when compared on the new EOS test simulations (Fig.~S2a).

To further assess the effects of potential data leakage using the original EOS and free-energy density difference test datasets, we also implemented an alternative split using a fragment-level sequence-identity filter.
Under this scheme, we ensured that no two fragments originating from the same full-length IDR and sharing 10 contiguous residues were split between the training and test sets.
This procedure yielded a filtered training dataset of 123,430 samples, while leaving the original test dataset unchanged.
We then retrained the MLP and PW models from scratch using this alternative filtered training set and evaluated their performances on the original test datasets.
Again, the test performances of these alternative retrained models are nearly indistinguishable from those of the original models (Fig.~S2b,c).

Together, these analyses establish that the predictive power of the learned MLP and PW models is minimally influenced by data leakage among the IDR fragments in the training and test sets.

\subsection{Comparison with mean-field models}

To compare the sequence-dependent pressure contributions predicted by mean-field pairwise models with simulation data, we remove contributions that depend only on total concentration.
Specifically, we subtract a fitted reference-pressure function from the simulated pressures, $\Delta_{\mathrm{PW}} P_{\mathrm{sim}} = P_{\mathrm{sim}}-\hat P^{\mathrm{ref}}(c_{\mathrm{tot}})$.
For the mean-field pairwise models, we likewise subtract the mean interaction contribution at fixed total concentration, $\Delta_{\mathrm{PW}} \hat P = (\nicefrac{1}{2})\sum_{i,j}\varepsilon_{ij}c_ic_j - (\nicefrac{1}{2})\bar\varepsilon\,(c_{\mathrm{tot}})^2$, where $\bar\varepsilon = \langle \sum_{ij}\varepsilon_{ij}y_iy_j\rangle$ is the mean over all mixtures in the training dataset, with $y_i=c_i/c_{\mathrm{tot}}$.

For FINCHES,\cite{ginell2025sequence} we used the unscaled $\tilde{\varepsilon}_{ij}$ coefficients returned by \texttt{calculate\_epsilon\_value}.
We selected the ``Mpipi\_original'' force field for comparisons with Mpipi simulations and the ``CALVADOS2'' force field for comparisons with CALVADOS simulations.
Temperature, pH, cutoff distance, and salt concentration were set to match the simulations, and the coefficients were normalized by chain length, $\varepsilon_{ij}=\tilde{\varepsilon}_{ij}/L$.

For Adachi et al.,\cite{adachi2024predicting} we computed the dimer-pair second virial coefficients $B^{\mathrm{DP}}_{ij}$ from their Eq.~(8).
Matching the second virial coefficient of the Flory--Huggins equation of state gives $\varepsilon_{ij}=2B^{\mathrm{DP}}_{ij}/(L^2v_0)-1$, an affine function of $B^{\mathrm{DP}}_{ij}$.
The additive constant cancels upon mean subtraction, and the multiplicative prefactor does not affect the Pearson correlation coefficient.
We therefore use $B^{\mathrm{DP}}_{ij}$ directly, defining $\Delta_{\mathrm{PW}} \hat P = \sum_{i,j}B^{\mathrm{DP}}_{ij}c_ic_j - \bar B\,c_{\mathrm{tot}}^2$, where $\bar B = \langle \sum_{ij}B^{\mathrm{DP}}_{ij}y_iy_j\rangle$ is the corresponding training-dataset mean.

The predictions of these mean-field pairwise models, along with the predictions of the learned pairwise model, are compared with simulation data in Fig.~S4.

\subsection{Prediction of multicomponent phase diagrams}

\subsubsection{Binary phase diagrams}

The binary phase diagrams shown in Figs.~4 and 5 were generated directly from the predicted free-energy densities using a multicomponent generalization of the standard common tangent construction.\cite{mao2019phase}
For a given two-component system, the binary composition space was first discretized into a square lattice, where $c_1$ and $c_2$ take values from 0 to \SI{0.25}{\chain\per\nm\cubed} with a spacing of \SI{0.005}{\chain\per\nm\cubed}.
Grid points with $c_{\mathrm{tot}}> \SI{0.25}{\chain\per\nm\cubed}$ were excluded from consideration.
The predicted free-energy densities, $\hat f$, of the binary mixtures corresponding to the resultant 1,326 grid points were then computed using \eqref{eq:fdecomposition}, with the integral term replaced by the excess free-energy density predicted by the model, $\hat{f}^{\mathrm{ex}}$.

The convex hull of the discretized free-energy density surface was then constructed using the algorithm developed by Mao et al.\cite{mao2019phase}
The constructed convex hull consists of triangular facets that envelop the discretized free-energy density surface from below.
Each facet comprises three vertices that define a common tangent plane to this surface.
Thus, the phases corresponding to the vertices of each facet satisfy the conditions of equal pressure and equal chemical potentials for all components, and are therefore in thermodynamic coexistence.

Each binary phase diagram was obtained by projecting the constructed convex hull back onto the composition space.
For each triangular facet, its vertices were first projected to define a triangle in the composition space.
If no free-energy grid point with composition lying inside this triangle was located above the corresponding common-tangent plane, no edges were drawn between the vertices, and the vertices were colored blue to denote a single-phase region.
Otherwise, edges were drawn between the projected vertices and colored orange to denote phase-separated regions.

In general, when a projected triangle contains only two elongated edges, it corresponds to predicted two-phase coexistence between mixtures with concentration vectors $c^\mathrm{I}$ and $c^{\mathrm{II}}$.
The elongated edges approximate the tie-lines connecting the two coexisting phases, while the two vertices joined by the short edge correspond to a single phase whose composition is approximately given by the average of the two vertices.
A mixture with a parent (i.e., overall) concentration vector lying within such a triangle is predicted to phase-separate into the two phases connected by the tie-line, with volume fractions determined by the lever rule.

When a projected triangle has three elongated edges, its three vertices denote the concentration vectors of three coexisting phases ($c^\mathrm{I}$, $c^{\mathrm{II}}$ and $c^{\mathrm{III}}$).
A mixture with a parent concentration vector $c$ that resides inside the triangle is predicted to phase-separate into these three phases with volume fractions $\eta^{\mathrm{I}}$, $\eta^{\mathrm{II}}$ and $\eta^{\mathrm{III}}$, such that $c=\sum_{\alpha}c^\alpha\eta^{\alpha}$ and $\sum_{\alpha}\eta^{\alpha}=1$ ($\eta^\alpha \ge0$).
Three-phase coexistence was, however, neither predicted nor observed in the binary systems chosen for phase diagram investigation.

\subsubsection{Ternary phase diagrams}

The ternary phase diagrams shown in Figs.~S7 and S8 were constructed using the same procedure as the binary phase diagrams described above, with the following modifications.
The three-dimensional composition space $(c_1,c_2,c_3)$ was discretized on a cubic lattice with the same spacing, and the free-energy density was evaluated at each lattice point.
We then constructed the lower convex hull of the discretized free-energy surface.
In this case, each simplicial convex-hull facet is a tetrahedron rather than a triangle.
Coplanar simplices were merged, while degenerate facets whose projections into composition space have zero volume were discarded.

With four vertices per tetrahedron, there are four possible phase classes.
We classified each projected tetrahedron according to its geometry, using a distance tolerance $\delta=2\Delta c=\SI{0.01}{\chain\per\nm\cubed}$, and the following hierarchical criteria:
\begin{itemize}\itemsep0ex
\item Single phase:
The longest edge is shorter than $\delta$, indicating that all four vertices coincide within the resolution of the lattice.
\item Two phases:
Otherwise, both vertices not belonging to the longest edge lie within a distance $\delta$ of the line containing that edge.
The tetrahedron thus collapses approximately onto a tie line connecting two coexisting phases.
\item Three phases:
Otherwise, the vertex opposite the largest-area triangular face lies within a distance $\delta$ of the plane containing that face.
The tetrahedron thus collapses approximately onto a tie triangle connecting three coexisting phases.
\item Four phases (not observed):
All remaining tetrahedra, whose four vertices span a nondegenerate volume in composition space, are classified as four-phase coexistence.
\end{itemize}

\subsection{Simulation of phase coexistence and comparison with predicted phase behavior}

To sample phase diagrams for testing, we partitioned the metric space uniformly into a 21-cell grid.
For the training/validation and held-out sequence sets separately, we selected a representative sequence from within a small hypersphere centered in each grid cell.
For each of these sequence sets, we constructed all unique sequence pairs, including homotypic systems, yielding 231 unary or binary systems.
We then used our trained model to predict the phase diagram for each system in each test set.

\subsubsection{Binary phase diagrams}

To validate the predicted phase behaviors, we randomly sampled sequence pairs and parent concentrations from the 231 representative phase diagrams in each test set for direct simulation.
To ensure balanced coverage of homogeneous and phase-separating mixtures, as well as homotypic (unary) and heterotypic (binary) phase separation, we stratified mixtures according to these categories based on the predicted phase diagrams and sampled separately from each stratum.
Within predicted two-phase regions, parent concentrations were chosen at the midpoints of the predicted tie lines.
To reduce ambiguous or slow-to-converge simulations, we excluded mixtures predicted to lie near phase boundaries.
In total, we obtained 59 test mixtures composed of sequences used in training and validation (Fig.~4), and 58 test mixtures composed of held-out sequences (Fig.~S6).

We performed direct-coexistence simulations of binary mixtures in a slab geometry using 1,000 chains with $L_z/L_x = L_z/L_y = 6$.
To verify equilibration, four simulations of each mixture were initialized from four distinct configurations: with components initially mixed or demixed, and with sequences initially distributed uniformly throughout the box or confined to half of the box.
Simulations were continued until the four realizations converged qualitatively.
Trajectories were recorded every 10{,}000 steps (\SI{100}{\ps}).
Local concentrations were computed by dividing the long axis $z$ into 50 bins.
For each test mixture, the realization exhibiting the smallest fluctuations in the concentration profile over the final 1,000 recorded frames was selected for downstream analysis.

For the selected realization of each simulated test mixture, we collected local concentrations of the two sequences, $(c^{\mathrm{local}}_1, c^{\mathrm{local}}_2)$, over the final 500 snapshots and estimated their joint distribution by binning samples into $\SI{0.005}{\chain\per\nm\cubed} \times \SI{0.005}{\chain\per\nm\cubed}$ bins.
This distribution is strongly peaked around the average concentrations of the coexisting bulk phases at equilibrium.
To compare model predictions with simulations, we converted the predicted equilibrium phases into a discretized probability distribution on the same concentration grid, assigning probability mass to each $(c_1,c_2)$ bin according to the predicted phase compositions and their corresponding volume fractions.
The predicted and simulated local concentration distributions were compared using the 1-Wasserstein (earth-mover) distance, computed with the \texttt{emd} method in the POT package.\cite{flamary2021pot}

To account for fluctuations and interfacial contributions to the concentration distributions in the simulations, we computed a best-case (positive-control) 1-Wasserstein distance defined as the minimum distance between the simulated distribution and any binned one-phase or two-phase distribution that preserves the parent concentration in the thermodynamic limit.
For two-phase distributions, we enforced a minimum Euclidean separation of \SI{0.05}{\chain\per\nm\cubed} between the concentrations of the two phases when constructing the positive control.
For each test set, we report the 1-Wasserstein distance between each model prediction and the corresponding simulated concentration distribution, normalized by the lower-bound distance of the positive control.
We also constructed a negative control for each test set by pairing simulations with random phase-coexistence predictions that preserve the parent concentration as well as the overall balance between one-phase and two-phase coexistence observed in the test simulations.

\subsubsection{Ternary phase diagrams}

For ternary mixtures, we selected two representative systems, each composed of three distinct sequences used in training and validation (Fig.~4a).
From each predicted ternary phase diagram, we selected parent concentrations in the predicted one-, two-, and three-phase regions, yielding nine test mixtures in total.
Direct-coexistence simulations were performed using 2,400 chains with $L_z/L_x = L_z/L_y = 8$.
To assess equilibration, each mixture was initialized from nine distinct configurations spanning different degrees of initial mixing and spatial segregation.
These included configurations in which all three components were mixed, all three were demixed, or one component was separated from the other two, with chains initially distributed throughout the simulation box or confined to one half of the box.
Although the simulations exhibited qualitative convergence, exceedingly long equilibration times prevented complete equilibration from all initial configurations.
The best-converged direct-coexistence simulations, judged based on stationarity and extent of fluctuations in the concentration profile, are compared with the MLP-model predictions in Figs.~S7 and S8.

\subsection{Prediction of multicomponent condensation}
\label{sec:condensation_prediction}

To predict multicomponent condensation directly from the metric space, we consider a scenario in which one or more high-total-concentration condensed phases coexist with a dilute (i.e., low-total-concentration) phase.
The presence of a stable dilute phase at thermodynamic equilibrium implies that the coexistence pressure is near zero, since the pressure in the dilute phase must be near zero according to the ideal gas law.
We therefore look for solutions of the equation
\begin{equation}
  \label{eq:P_eq_zero}
P(c_{\mathrm{tot}}; \bar z) = c_{\mathrm{tot}} kT + P^{\mathrm{ex}}(c_{\mathrm{tot}} \bar z) \approx 0,
\end{equation}
where $c_{\mathrm{tot}} \ge 0$ is the total concentration and $\bar z \in \mathbb{R}^d$ denotes a concentration-weighted average point in the metric space.
We emphasize that $\bar z$ is related to the putative condensed phase, not the parent concentration of a phase-separated mixture.
Only solutions satisfying
\begin{equation}
  \label{eq:P_eq_zero_stable}
\frac{dP}{dc_{\mathrm{tot}}} > 0
\end{equation}
are retained, corresponding to mechanical stability with respect to total-concentration fluctuations at fixed mixture composition.
We ignore the trivial solution at $c_{\mathrm{tot}} = 0$, which corresponds to the dilute phase in this $P \approx 0$ approximation.

By determining whether nontrivial solutions to \eqref{eq:P_eq_zero} and \eqref{eq:P_eq_zero_stable} exist at a particular value of $\bar z$, we can immediately conclude whether a condensed phase with mixture representation $\bar z$ can coexist with a dilute phase within the $P \approx 0$ approximation.
Importantly, this solution does not specify how $\bar z$ is constructed: it can represent the feature vector of a single sequence, or it can represent the concentration-weighted average of the feature vectors of many distinct sequences in a mixture, in which case $\bar z$ must lie within the convex hull of the feature vectors of those sequences.

We note that the existence of a solution to \eqref{eq:P_eq_zero} and \eqref{eq:P_eq_zero_stable} does not necessarily imply that a condensed phase with mixture representation $\bar z$ will be stable at global thermodynamic equilibrium in a multicomponent mixture, since there may in general be competing phases whose free energies must also be considered.
Moreover, we note that this approach does not predict how many distinct condensed phases are in coexistence with the dilute phase.
Finally, we emphasize that this construction only determines whether a condensed phase can coexist with a dilute phase; it does not indicate whether other types of phase-separated states, specifically those in which none of the coexisting phases are dilute, are possible.

In practice, we solved \eqref{eq:P_eq_zero} for $c_{\mathrm{tot}} \in (0,c_{\mathrm{tot},\max}=\SI{0.25}{\chain\per\nm\cubed}]$ by bracketing roots on a uniform grid in $c_{\mathrm{tot}}$ and refining candidate solutions using a scalar root-finding algorithm.
We retained the solution with the lowest $c_{\mathrm{tot}}$ if multiple solutions were found.
For visualization purposes, we performed these calculations on a two-dimensional lattice in $\bar z$, which was selected by taking a slice of the full $d$-dimensional metric space.
We then identified both an example single sequence $s_i$ whose feature vector lies within $\sim\mathrm{RMSD}_i$ of a lattice point with a solution to \eqref{eq:P_eq_zero}, and a pair of sequences $(s_i, s_j)$ for which the line segment between their feature vectors passes within $\sim\max(\mathrm{RMSD}_i,\mathrm{RMSD}_j)$ of a lattice point with a solution to \eqref{eq:P_eq_zero}.

\subsection{Analysis of point mutation distances}
\label{sec:point_mutation_distances}

We used the metric property of our model to compute the average effect size of single substitutions from amino-acid type $a \in \mathcal{A}$ to amino-acid type $b \in \mathcal{A}$.
For each WT sequence in the test set, let $z_{\mathrm{WT}}$ denote the feature vector of the WT sequence and let $z_m$ denote the feature vector of a single-substitution mutant indexed by $m \in \{1,\ldots,19L\}$ derived from that WT sequence.

Let $\mathcal{M}(a\!\to\!b)$ be the set of all single-substitution mutants derived from a WT sequence whose substitution is $a\!\to\!b$.
To compute the average effect size, we first compute the PLL-weighted average displacement in the metric space for each substitution type given a WT sequence,
\begin{equation}
  \Delta z(a\!\to\!b) = 
\frac{\sum\limits_{m \in \mathcal{M}(a\to b)} w_m \, \left| z_m - z_{\mathrm{WT}} \right|}
{\sum\limits_{m \in \mathcal{M}(a\to b)} w_m},
\end{equation}
with weights $w_m \propto \exp (\mathrm{PLL}_m)$.
We then aggregate these results over all WT sequences in the test set, indexed by $i = 1,\ldots,n$, to construct the matrix
\begin{equation}
\left\langle \left| z_{\mathrm{MUT}} - z_{\mathrm{WT}} \right| \right\rangle_{a\to b}
\;=\;
\frac{1}{|\mathcal{R}_{a\to b}|}
\sum_{i \in \mathcal{R}_{a\to b}} \Delta z_i(a\!\to\!b),
\end{equation}
where $\mathcal{R}_{a\to b}$ denotes the WT indices for which $\mathcal{M}(a\!\to\!b)$ is nonempty.
For visualization and comparison, we sort substitution types by the dominant left singular mode of $\langle \left| z_{\mathrm{MUT}} - z_{\mathrm{WT}} \right| \rangle_{a\to b}$, which orders WT amino-acid types by the principal pattern of variation in their average mutation-induced displacements.

\subsection{Theoretical sequence descriptors}
\label{sec:descriptors}

We constructed a set of 60 theoretical descriptors defined in the IDR literature (listed in Table~\ref{tab:descriptors}) to characterize the compositional and patterning properties of the IDR fragments studied in this work.

\clearpage

\newcommand{\WrapDesc}[1]{\parbox[t]{0.82\linewidth}{\raggedright #1}}
\newcommand{\WrapFeat}[1]{\parbox[t]{2.2cm}{\raggedleft #1}}
{
\setlength{\LTcapwidth}{\linewidth}
\newcommand{\RowGap}{\noalign{\vskip 10pt}}

\begin{longtable}{@{}rcl@{}}
\renewcommand{\arraystretch}{1.5}

\\
\hline
\textbf{Descriptor} & & \textbf{Description} \\
\hline\noalign{\vskip 6pt}

NCPR &
&
\WrapDesc{Net charge per residue is defined as NCPR=$\frac{1}{L}\sum_{l=1}^L q_l$, where $q_l$ is the charge of the residue at sequence position $l$ and $L$ is sequence length. Residue charges were assigned according to the Mpipi force field,\cite{joseph2021physics} in which R and K carry charges of $+0.75\,e$, H carries $+0.35\,e$, D and E carry $-0.75\,e$, and all other residues are charge-neutral.}
\\
\RowGap

FCR &
&
\WrapDesc{Fraction of charged residues, FCR, of a sequence is defined as its number of charged amino acids divided by its total number of amino acids.}
\\
\RowGap

SCD &
&
\WrapDesc{Sequence charge decoration is defined as $\mathrm{SCD}=\frac{1}{L}\left[\sum_{l=2}^{L}\sum_{m=1}^{l-1} q_l q_m (l-m)^{1/2}\right]$,\cite{sawle2015theoretical_SCD} where $q_l$ and $q_m$ denote the charges of residues at sequence positions $l$ and $m$, respectively, and $L$ is the sequence length.}
\\
\RowGap

$\kappa$ &
&
\WrapDesc{The parameter $\kappa$ characterizes the linear sequence distribution of oppositely charged residues \cite{das2013conformations} and was computed using localCIDER.\cite{holehouse2017cider} Low and high values of $\kappa$ correspond to well-mixed and segregated charge distributions, respectively.}
\\
\RowGap

$\left<\mathrm{size}\right >$ & 
&
\WrapDesc{In the Mpipi force field, each amino acid is associated with 20 $\sigma$ parameters that characterize the repulsive core length scales of its nonbonded interactions with each of the 20 amino acid types, including itself (see \secref{sec:mpipiFF} and \eqref{eq:Mpipi_potential_WF}). For each amino acid, we define $\bar{\sigma}$ as the average of its 20 $\sigma$ values. The average residue size of a sequence is computed as $\left<\mathrm{size}\right >=\frac{1}{L}\sum_{l=1}^L\bar{\sigma_l}$, where $\bar{\sigma}_l$ is the $\bar{\sigma}$ of the amino acid at sequence position $l$ and $L$ is sequence length.}
\\
\RowGap

$\left<\mathrm{hydrophob.}\right >$ &
&
\WrapDesc{In the Mpipi force field, each amino acid is associated with 20 $\varepsilon$ parameters that scale its nonbonded interaction potential with each of the 20 amino acid types, including itself (see \secref{sec:mpipiFF} and \eqref{eq:Mpipi_potential_WF}). For each amino acid, we define $\bar{\varepsilon}$ as the average of its 20 $\varepsilon$ values. The average residue hydrophobicity is computed as $\left<\mathrm{hydrophob.}\right >=\frac{1}{L}\sum_{l=1}^L\bar{\varepsilon_l}$, where $\bar{\varepsilon}_l$ is the $\bar{\varepsilon}$ of the amino acid at sequence position $l$ and $L$ is sequence length.}
\\
\RowGap

$\mathrm{WF}_{\mathrm{pairs}}$ &
&
\WrapDesc{The descriptor $\mathrm{WF}_{\mathrm{pairs}}$ is the average of the scaled integral of the attractive part of the nonbonded potential over all pairs of residues in a sequence, given by $\mathrm{WF}_{\mathrm{pairs}}=\frac{1}{L(L+1)/2}\sum_{l=1}^{L}\sum_{m=l}^L\int_{\sigma_{lm}}^{3\sigma_{lm}}4\pi r^2\phi_{lm}(r)dr$, where $\sigma_{lm}$ and $\phi_{lm}$ are the parameters of the nonbonded potential between the residues at sequence positions $l$ and $m$ (see \secref{sec:mpipiFF}) and $L$ is sequence length.}
\\
\RowGap

SHD &
&
\WrapDesc{Sequence hydropathy decoration SHD is computed as SHD=$\frac{1}{L}\sum_{l=1}^L\sum_{m=l+1}^L\varepsilon_{lm}|m-l|^{-1}$,\cite{zheng2020hydropathy} where $\varepsilon_{lm}$ scales the nonbonded interaction potential between residues at sequence positions $l$ and $m$ and $L$ is sequence length.}
\\
\RowGap

\WrapFeat{Composition descriptors}&
&
\WrapDesc{We computed the fractions of residues in the following twelve residue groups to define twelve amino-acid composition descriptors: \{G\}, \{D,E\} ``negatively charged'', \{A,C,I,L,M,P,S,T,V\} ``neutral'', \{N,Q\} ``neutral with $\pi$ electron'', \{K\} ``positively charged'', \{H,R\} ``positively charged with $\pi$ electron'', \{F,W,Y\} ``aromatic'', \{K,R\}, \{K,R,H\}, \{W\}, \{F\}, and \{Y\}. The first seven groups are defined in Ref.~\cite{joseph2021physics}. The last five groups were included to investigate the distinct roles played by different aromatic and positively charged residues in the Mpipi force field. Each of these composition descriptors is labeled as ``Fraction X'', where X is the residue subset (e.g., Fraction DE).}
\\
\RowGap

\WrapFeat{Mpipi composition}&
&
\WrapDesc{The twelve composition descriptors described above were concatenated to construct the Mpipi composition descriptor set.}
\\
\RowGap

\WrapFeat{NARDINI descriptors}&
&
\WrapDesc{We used NARDINI descriptors to characterize residue patterning for the first nine residue groups defined in the composition descriptors above.\cite{cohan2022uncovering} The $\Omega$ descriptor for a residue group X quantifies the binary patterning of residues either belonging to or excluded from the specified group and is labeled as $z(\Omega_{\mathrm{X}})$ (e.g., $z(\Omega_{\mathrm{D}\mathrm{E}})$). The $\delta$ descriptor for a pair of residue groups X and Y quantifies the binary patterning of residues belonging to the two groups and is labeled as $z(\delta_{\mathrm{X,Y}})$ (e.g., $z(\delta_{\mathrm{DE},\mathrm{KR}})$). We excluded pairs of groups with overlapping residues in the computation of the $\delta$ descriptors, resulting in a total of forty NARDINI descriptors (nine $\Omega$-family descriptors and thirty-one $\delta$-family descriptors).}
\\
\RowGap

\WrapFeat{NARDINI}&
&
\WrapDesc{The forty NARDINI descriptors described above were concatenated to construct the NARDINI descriptor set.}
\\

\noalign{\vskip 6pt}
\hline
\noalign{\vskip 10.0pt}

\caption{Theoretical sequence descriptors and descriptor sets used to characterize compositional and patterning properties of IDR sequences.}
\label{tab:descriptors}

\end{longtable}
}

\subsection{Explained variance and canonical correlation analysis of theoretical descriptors}
\label{sec:explained_variance}

We analyzed linear relationships between the metric representation and the theoretical descriptors defined in \secref{sec:descriptors}.
The feature vectors $z_i \in \mathbb{R}^d$ for IDRs $i = 1,\dots,S$ were stacked to form the matrix $Z \in \mathbb{R}^{S \times d}$.
Corresponding $D$-dimensional theoretical descriptor vectors, denoted $y_i \in \mathbb{R}^D$, were stacked to form the matrix $Y \in \mathbb{R}^{S \times D}$.
Because the feature vectors represent coordinates in a metric space with a physically meaningful origin, they were neither centered nor rescaled in these analyses.
By contrast, the theoretical descriptors were centered and variance-normalized as described below.

\paragraph{Per-descriptor explained variance and metric-space loadings.}
We first centered and normalized each individual theoretical descriptor $Y_r \in \mathbb{R}$, yielding a whitened scalar descriptor with zero mean and unit variance.
We then computed the fraction of the variance of the whitened descriptor $Y_r$ explained by the metric representation $Z$ via a no-intercept ordinary least squares projection onto the span of $Z$,
\begin{equation}
R^2(Y_r \mid Z) \;=\; \frac{\lVert \hat y_r \rVert^2}{\lVert y_r \rVert^2},
\end{equation}
where $\hat y_r$ denotes the projection of the whitened theoretical descriptor onto the metric space.
This quantity measures how well the feature vectors explain variability in each theoretical descriptor.
We also computed the descriptor-aligned metric-space loading vector $v_r \in \mathbb{R}^d$.
We visualized the squared components $v_{rj}^2$, which quantify the contribution of each metric-space dimension $j$ to descriptor $r$; the squared components are independent of the sign of the loading and satisfy $\sum_{j=1}^d v_{rj}^2 = 1$.

\paragraph{Multivariate explained variance for descriptor subsets and metric-space contributions.}
To assess how well the metric representation explains collections of multiple theoretical descriptors, we computed a multivariate coefficient of determination $R^2(Y \mid Z)$ for selected $D$-dimensional subsets of descriptors.
For each selected descriptor subset, we centered and whitened the columns of $Y \in \mathbb{R}^D$ to remove scale and correlation effects.
We then projected the whitened descriptor matrix onto the span of $Z$ using no-intercept linear regression and computed the explained variance
\begin{equation}
R^2(Y \mid Z)
\;=\;
\frac{\operatorname{tr}\!\big(\operatorname{Cov}(\hat Y)\big)}
     {\operatorname{tr}\!\big(\operatorname{Cov}(Y)\big)},
\end{equation}
where $\hat Y$ denotes the projection of the whitened theoretical descriptors onto the metric space.
This quantity reports the total fraction of variance of this collection of theoretical descriptors explained by the feature vectors.

For each selected descriptor subset, we used canonical correlation analysis between $Z$ and the whitened descriptor matrix to assess the contribution of each metric-space dimension to the explained variance.
We first computed the squared canonical correlations $\{\rho_k^2\}$ and corresponding metric-space canonical loading vectors $\{v_k \in \mathbb{R}^d\}$, where $k$ indexes the canonical modes.
We then summarized how strongly each metric-space dimension contributes across all canonical modes by defining a $\rho^2$-weighted average of the squared metric-space canonical loading components,
\begin{equation}
  \bar v_j^2
  \;=\;
  \frac{\sum_k \rho_k^{2}\, v_{kj}^{2}}
       {\sum_k \rho_k^{2}},
       \qquad j = 1,\dots,d.
\end{equation}
This quantity measures the overall importance of metric-space dimension $j$ in explaining the variance across selected theoretical descriptors by emphasizing canonical modes with stronger feature vector--descriptor coupling.
As in the single-descriptor case, the squared components are independent of the sign of the loading and satisfy $\sum_{j=1}^d \bar v_{j}^2 = 1$.

\subsection{Mutual information analysis}

Mutual information between a descriptor (or descriptor set) $Y$ and the metric-space representation $Z$ was estimated using the SMILE method.\cite{song2019smile} For each of the 335,439 sequences in our curated IDR fragment ensemble (see \secref{sec:idr_curation}), the corresponding value of $Y$ was paired with metric-space representations $Z$ generated by each of the five encoders trained in five-fold cross-validation (see \secref{sec:train_val_test}), resulting in 1,677,195 joint ($Y$,$Z$) samples.

The 235,439 training and validation sequences were randomly divided into three partitions for mutual information evaluation.
For each fold $k=1,2,3$, two consecutive partitions ($k$ to $k+1$, cyclically) were used for training, the remaining partition for validation, and the 100,000 held-out test sequences for testing.
Mutual information $I(Y,Z)$ was estimated with the joint $(Y,Z)$ data of each fold using the \texttt{estimate\_mi} function from the \texttt{torch\_mist} package,\cite{torch_mist} with a patience of 10 epochs, a $\tau$ value of 25, and a critic neural network consisting of two hidden layers with 256 neurons in each.
This procedure produced three test estimates of $I(Y,Z)$, whose average was recorded.
The entire process was repeated five times, yielding a total of five average test estimates.
The mean and standard deviation of these five estimates are reported as the value of $I(Y,Z)$ and its uncertainty, respectively.

The entropy of the metric-space representation, $H(Z)$, was estimated by computing $I(Z_{\mathrm{avg}},Z)$ using the same procedure as for $I(Y,Z)$, where $Z_{\mathrm{avg}}$ denotes the model-averaged feature vectors obtained from the five transformer encoders trained via five-fold cross-validation.
In the \texttt{estimate\_mi} function, a batch size of 4,096 was used when evaluating $I(Y,Z)$ for all the 60 individual descriptors (except for $z(\delta_{\mathrm{K,FWY}})$) and the descriptor set NARDINI (see \secref{sec:descriptors}), while a batch size of 65,536 was used in all other cases for stability.

\subsection{Counterfactual analyses using mutant fragments}
\label{sec:counterfactual_analyses}

To isolate sequence-dependent effects, we analyzed the thermodynamic consequences of mutations that preserve amino-acid composition.
Following the approach described in \secref{sec:sequence_generation}, we constructed all unique composition-preserving single-swap mutant sequences.
We refer to these mutants as counterfactuals, since any change in their metric-space representation relative to the corresponding WT sequence can be attributed to the ordering of the amino acids within the sequence rather than amino-acid identity.





We then analyzed (a)~the overall distribution of metric-space distances between these mutants and their respective WT sequences, (b)~the position dependence of mutational effects within the sequences, and (c)~the influence of short motifs within the WT sequences on these mutational effects.
In all cases, we computed weighted averages over all sequences derived from a WT sequence with weights $w_m \propto \exp(\mathrm{PLL}_m)$ and then aggregated these results over all WT sequences in the 100,000-sequence test set (see \secref{sec:sequence_generation}).

\paragraph{Global distribution of counterfactual spread.}
For each WT sequence $i$ in the test set, we computed the spread of metric-space distances over the set $\mathcal{S}_i$ consisting of the WT sequence and all associated unique single-swap mutants,
\begin{equation}
\mathrm{Spread}_i =
\left\langle
\left\|
z_m -
z_i^{\mathrm{cent}}
\right\|_2^2
\right\rangle_i^{1/2},
\qquad
z_i^{\mathrm{cent}} = \langle z_m \rangle_i ,
\end{equation}
where $m \in \mathcal{S}_i$ and the averages are PLL-weighted.
To summarize counterfactual variability across the test set, we aggregated the spreads associated with all WT sequences into an ensemble-level histogram.
We then analyzed typical properties of the WT sequences whose mutants lead to the highest 10\% and lowest 10\% spreads in the ensemble-level histogram.
To this end, we computed histograms of individual whitened theoretical descriptors using WT sequences in either the top-10\% or bottom-10\% of the spread distribution.
We then ranked the relevance of these descriptors for distinguishing these two classes of WT sequences by computing the empirical mutual information between the class label and the pair of histograms for each whitened theoretical descriptor.

\paragraph{Position-dependent counterfactual spread.}
To assess positional dependence, we restricted $\mathcal{S}_i$ to the WT sequence and the subset of unique single-swap mutants whose sequence differs from the WT at position $l$.
For each position $l \in \{1,\dots,L\}$, we computed the corresponding spread $\mathrm{Spread}_{i,l}$ using the same definition as above.
We then aggregated these values over WT sequences to obtain the ensemble-level, position-dependent
spread
\begin{equation}
\label{eq:spread_ens_average}
\langle \mathrm{Spread}_l \rangle =
\frac{1}{|\mathcal{R}_l|}
\sum_{i \in \mathcal{R}_l}
\mathrm{Spread}_{i,l},
\end{equation}
where $\mathcal{R}_l$ denotes the set of WT sequences for which position $l$ admits at least one single-swap mutant.
We similarly defined a position-dependent spread conditioned on the WT amino acid at the swapped position $l$ having type $a \in \mathcal{A}$, $\langle \mathrm{Spread}_{l,a} \rangle$, by further restricting the set $\mathcal{S}_i$ and aggregating the results analogously to \eqref{eq:spread_ens_average}.

\paragraph{Motif-level analysis of counterfactual spread.}
Finally, we analyzed counterfactual effects associated with short sequence motifs defined as ordered tuples of residue classes, $M = (C_1,\dots,C_{L_{\mathrm{motif}}})$, where each wildcard $C_\ell \subset \mathcal{A}$ specifies an allowed set of amino acids and $L_{\mathrm{motif}}$ is the motif length.
The wildcards utilized in this work are presented in \tabref{tab:motif_wildcards}.
Motifs are treated as equivalent under reversal, defining a canonical motif class $[M] = \{M, M^{\mathrm{rev}}\}$.
A WT sequence is said to satisfy a motif if it contains a contiguous subsequence whose residues match the wildcard classes in $M$ or its reversed counterpart $M^{\mathrm{rev}}$.
A corresponding motif-control class is defined by requiring only that the same wildcard classes be satisfiable at distinct positions anywhere in the WT sequence, in any order, without positional contiguity.
For a given WT sequence $i$, we formed the set $\mathcal{S}_i$ consisting of the WT sequence and all unique single-swap mutants that alter at least one WT position participating in a satisfying match of the motif or its control.
For each motif and control class, we computed the corresponding spread over $\mathcal{S}_i$ as above and then aggregated these values across WT sequences following the same logic as \eqref{eq:spread_ens_average}.

\begin{table}[t]
\centering
\begin{tabular}{lcl}
\hline
Wildcard class$\;\;$ & $\;\;$Symbol$\;\;$ & Amino acids \\
\hline\noalign{\vskip 6pt}
Positive & $+$ & K, R, H \\
Negative & $-$ & D, E \\
Hydrophobic & h & A, F, I, L, M, V, W, Y \\
Polar & p & C, N, Q, S, T \\
Any & $*$ & A, C, D, E, F, G, H, I, K, L, M, N, P, Q, R, S, T, V, W, Y \\
\noalign{\vskip 6pt}\hline
\end{tabular}
\caption{Definitions of amino-acid wildcard classes used in motif specifications.}
  \label{tab:motif_wildcards}
\end{table}

\clearpage
\section{Datasets and computational resources used}
\label{sec:resource}
\begin{table}[h!]
\vbox{
{\centering
{\small
\begin{ruledtabular}
  \begin{tabular}{lccccc}
\makecell{Type of\\ simulation} &
\makecell{CPU core-\\hours per sim.} &
\makecell{GPU hours\\ per sim.} &
\makecell{Dataset\\ name} &
\makecell{Number of\\elements} &
\makecell{Total\\computation\\resources \\(CPU core-hours; \\GPU hours)} \\
\hline
\multirow{16}{*}{EOS} &
\multirow{16}{*}{4--12} &
\multirow{16}{*}{n/a} &
EOS$\_$unilat$\_$test$\_$single & 10,050 & $\sim80,400$ \\
& & & EOS$\_$unilat$\_$test$\_$pair & 10,440 & $\sim83,520$ \\
& & & $\ldots$test$\_$pair$\_$MUT & 10,440 & $\sim83,520$ \\
& & & $\ldots$test$\_$triple & 10,050 & $\sim80,400$ \\
& & & $\ldots$test$\_$quad & 10,050 & $\sim80,400$ \\
& & & EOS$\_$resplit$\_$test$\_$single & 1,000 & $\sim$8,000 \\
& & & $\ldots$test$\_$pair & 1,000 & $\sim$8,000 \\
& & & $\ldots$test$\_$triple & 1,000 & $\sim$8,000 \\
& & & $\ldots$test$\_$quad & 1,000 & $\sim$8,000 \\
& & & EOS$\_$uniseq$\_$test$\_$single & 7,200 & $\sim57,600$ \\
& & & $\ldots$test$\_$pair & 7,140 & $\sim57,120$ \\
& & & $\ldots$test$\_$triple & 7,140 & $\sim57,120$ \\
& & & $\ldots$test$\_$quad & 7,140 & $\sim57,120$ \\
& & & EOS$\_$uniseq$\_$train$\_$pair & 100,000 & $\sim800,000$ \\
& & & $\ldots$train$\_$single & 20,000 & $\sim160,000$ \\
& & & $\ldots$train$\_$pair$\_$high$\_$conc & 20,000 & $\sim160,000$ \\

\hline
\multirow{2}{*}{\makecell{EOS\\CALVADOS}} &
\multirow{2}{*}{2--3} &
\multirow{2}{*}{0.5--0.75} &
EOS$\_$unilat$\_$test$\_$pair & 10,440 & $\sim27,100$; 6,800 \\
& & & $\ldots$test$\_$pair$\_$MUT & 10,440 & $\sim27,100$; 6,800 \\

\hline
\multirow{5}{*}{FE diff.} &
\multirow{5}{*}{\makecell{320--960\\(single comp.)\\640--1,920\\(two or more\\components)}} &
\multirow{5}{*}{n/a} &
FED$\_$unilat$\_$test$\_$single & 63 & $\sim40,320$ \\
& & & $\ldots$test$\_$pair & 231 & $\sim295,680$ \\
& & & $\ldots$test$\_$triple & 60 & $\sim76,800$ \\
& & & $\ldots$test$\_$quad & 60 & $\sim76,800$ \\
& & & FED$\_$uniseq$\_$test$\_$pair & 100 & $\sim128,000$ \\

\hline
\multirow{3}{*}{\makecell{Direct-\\coex.}} &
\multirow{2}{*}{10,752--13,824} &
\multirow{2}{*}{672--864} &
DC$\_$binary$\_$seen & 59 & $\sim724,992$; 45,312 \\
& & & DC$\_$binary$\_$unseen & 58 & $\sim712,704$; 44,544 \\
& 23,808--29,952 & 1,488--1,872 & DC$\_$ternary$\_$seen & 9 & $\sim 240,384$; 15,024  \\
\end{tabular}

\caption{Summary of computational resources required for EOS, free-energy (FE) difference, and direct-coexistence simulations. EOS datasets whose names contain train/test were used as training and validation or test data, respectively. The FED datasets were used to construct free-energy density difference test data via explicit thermodynamic integration. For both the EOS and FED datasets, the labels unilat/uniseq indicate that the IDR components in the corresponding systems were sampled uniformly from the $d=10$ MLP-model metric space or sequence space, respectively. The DC$\_$seen/DC$\_$unseen datasets were used to evaluate phase diagram predictions for test mixtures composed of IDR sequences that were either encountered (seen) or held-out (unseen) during EOS training and validation; in both cases, the test mixtures themselves were not encountered during EOS training and validation.}

\end{ruledtabular}
}}}

\end{table}

\clearpage

\addtocontents{toc}{%
  \protect\contentsline{section}%
  {\protect\numberline{IV}Supplemental Figures}%
  {29}{}
}